\documentclass[12pt]{article}

\usepackage{amssymb,amsmath,amsfonts,eurosym,float,geometry,ulem,graphicx,caption,color,setspace,comment,footmisc,caption,natbib,pdflscape,array,epstopdf,booktabs}
\usepackage{lscape} 
\usepackage{silence}
\usepackage{xr-hyper}
\makeatletter
\newcommand*{\addFileDependency}[1]{
  \typeout{(#1)}
  \@addtofilelist{#1}
  \IfFileExists{#1}{}{\typeout{No file #1.}}
}
\makeatother

\newcommand*{\myexternaldocument}[1]{
    \externaldocument{#1}
    \addFileDependency{#1.tex}
    \addFileDependency{#1.aux}
}
\myexternaldocument{online_appendix}

\usepackage[colorlinks=true,linkcolor=blue,citecolor=blue]{hyperref}
\usepackage{multibib}
\newcites{Appendix}{Appendix References}
\begin{document}

\begin{titlepage}
    \title{Smartphone Data Reveal Neighborhood-Level Racial Disparities in Police Presence} 
    \author{M. Keith Chen, Katherine L. Christensen, Elicia John, \\ Emily Owens, Yilin Zhuo\thanks{Chen: UCLA Anderson School of Management, Los Angeles, CA 90095, USA, keith.chen@naderson.ucla.edu. 
    Christensen: Indiana University, Kelley School of Business, Bloomington, IN 47405, USA, kachris@iu.edu. John: American University, Kogod School of Business, Washington, DC 20016, USA, eliciaj@american.edu. Owens: Department of Criminology, Law \& Society, Department of Economics, University of California Irvine, Irvine, CA 92697, USA, egowens@uci.edu. Zhuo: UCLA Anderson School of Management, Los Angeles, CA 90095, USA, yilin.zhuo.phd@anderson.ucla.edu. Acknowledgements: We thank Milind Rajavasireddy for excellent research assistance, as well as Bocar Ba, Kitt Carpenter, Amanda Geller, Michael Norton, Broderick Turner, and seminar participants in the Virtual Crime Economics (ViCE) Seminar and the USC-UMich-UVA Online Seminar on Law and Economics for helpful comments.}}
    \date{\today}
    \maketitle
        \begin{abstract}
    While extensive, research on policing in America has focused on documented actions such as stops and arrests—less is known about patrolling and presence. We map the movements of over ten thousand police officers across twenty-one of America’s largest cities by combining anonymized smartphone data with station and precinct boundaries. Police spend considerably more time in Black neighborhoods, a disparity which persists after controlling for density, socioeconomics, and crime-driven demand for policing. Our results suggest that roughly half of observed racial disparities in arrests are associated with this exposure disparity, which is lower in cities with more supervisor (but not officer) diversity. 
    \end{abstract}
    \setcounter{page}{0}
    \thispagestyle{empty}
    \end{titlepage}
    \pagebreak \newpage
    
\doublespacing

According to FBI statistics, in 2019 Black people in America were arrested at twice the rate of White people (\citealt{OJJDP}). A large literature explores the causes of racial disparities in policing outcomes such as stops, searches, and arrests, including differences in socioeconomic status, criminal activity, and biased decision making by police officers (\citealt{banaji2021systemic,banks2006discrimination, Hoekstra_Sloan2020, rucker2021toward}). We document and explore another potential factor: racial disparities in police presence.

The police wield considerable discretion in determining where, exactly, law enforcement is provided within America's cities. Where officers are located has direct implications for the deterrence of potential criminals. It also affects what the broader public knows about crime, as police presence can influence when and where crime is officially observed and recorded. Finally, since police contact is the first part of any criminal justice involvement, and the racial composition of neighborhoods varies both across and within cities, detailed information on where officers work during their shifts can potentially identify sources of disparities in later criminal justice outcomes.

Unfortunately, few departments collect detailed officer location data, and even fewer release it to researchers in a standardized way. We use anonymized smartphone location data to identify and track the movements of individual police officers while on patrol. These data identify where police spend their time, and allow us to evaluate spatial patterns of policing at scale while protecting officer privacy. Using these data, we quantify how patterns of socioeconomic status, crime, social capital, and race relate to local police “presence” within and across twenty-one of America’s largest cities. 

Generally, geographic analysis of policing at the sub-city level has measured local policing in two ways: “up-funnel” or “down-funnel” measurement. Like \cite{Ba_al2021b, vomfell2021officer,mastrobuoni2019police} and \cite{weisburd2021police}, we use an “up-funnel” measurement approach, capturing information on where police conduct patrol.  In contrast, “down-funnel” measures capture observed police enforcement actions that are made public by a department (e.g. use of force, arrests, or stops). We are able to build on existing up-funnel studies, as we can track exact officer location, rather than where officers are instructed to go.

We measure police presence as the total number of officer hours spent in a census block group (a “neighborhood” with roughly 1,000 residents) over a ten-month period (Feb 2017 - Nov 2017), when the officer was moving through a neighborhood at 50 mph or less. Comparisons of assigned officer beats, an alternative up-funnel measure used in, for example, \cite{Ba_al2021b}, with actual patrol car locations suggest that while beats are highly correlated with actual officer presence, officers spend a nontrivial amount of time outside of their officially assigned geographic locations (\citealt{weisburd2021police}). Further, unlike studies based on Automatic Vehicle Location (AVL) technology, we are able to capture the time patrol officers spend patrolling outside of their cars. Finally, we are able to extend beyond a single-city analysis based on Automatic Vehicle Location (AVL) or assigned beat data to a much broader set of US cities by using smartphone GPS data that are broadly representative of the United States population (\citealt{long2020political}). 

Our sub-city analysis of GPS location data confirms many hypotheses posed by qualitative and historical research on up-funnel police presence across America (\citealt{Hinton2016, Rios2011, Sharkey2018}), and patterns observed at the city level (\citealt{CarmichaelKent2014}). For example, we show that police officers spend more time in places with larger Black, Hispanic, or Asian populations relative to the city overall. While controlling for neighborhood differences in socioeconomic status, social disorganization, and violent crime reduces these disparities, it does not eliminate them. This suggests that social interventions targeted at the “root causes” of crime may be unlikely to eliminate the racial disparities we observe in American policing.  

Our up-funnel approach has distinct advantages over down-funnel approaches; enforcement-based measures of policing do not fully measure officer presence in a neighborhood - the most fundamental element of what it means for an area to be “policed,” and in particular fail to capture the extent to which police officers allocate time to the increasingly diverse set of tasks they perform within a community (\citealt{Brooks2021, Lum2021, Lumetal2021}). At the same time, our measure of officer presence complements existing spatial analyses of down-funnel, enforcement-based measures of policing by contributing analysis of pre-enforcement exposure to police at a highly granular level. For example, many studies have found that police engage in more enforcement actions in Black neighborhoods, with more mixed evidence in Hispanic neighborhoods (\citealt{Geller_al2014, Ba_al2021b, pierson2020large}). Combining our police presence data with geocoded arrest data that are available for six cities (including New York City), our estimates suggest that in this sub-sample, differences in where officers spend time explain roughly 60\% of the Black-White disparity, and almost 70\% of the Hispanic-White disparity in arrests. Officers' higher propensity to make an arrest, conditional on being in a non-White neighborhood, explains the remainder of these disparities. 

While still descriptive, we also explore differences in police presence associated with the racial composition of front-line officers versus police supervisors across cities. Consistent with existing single-department studies (e.g. \citealt{Hoekstra_Sloan2020, Ba_al2021b}), our results suggest that the additional police presence in Black neighborhoods is lower in cities where more police officers are Black.  However, this is not statistically explained by the racial composition of patrol officers walking the beat, but rather by the racial composition of the supervisors that direct patrol officer activity.  While not causal, this highlights the potential role of retention and promotion in police reform aimed at reducing racial disparities in the criminal justice system. We then provide evidence that the nature of disparities in policing presence differs across US cities. While some disparities in some cities (e.g. Charlotte, NC) are largely associated with spatial differences in socioeconomic status (e.g., income, education, and civic engagement) others persist even when controlling for these factors, including spatial patterns of violent crime (e.g. Austin, TX). 

Our findings suggest that disparities in exposure to police are associated with both structural socioeconomic disparities and discretionary decision making by police commanders and officers. Hence, this study provides novel data on police-civilian interaction to enable additional analyses of the factors driving these observed effects in hopes of developing policy interventions to mitigate them. Finally, our police presence measure provides a new benchmark against which down-funnel police actions like stops and arrests may be objectively evaluated.

\setcounter{section}{1}
\section{Methods}

\subsection{Data}

The smartphone location data used in this study were provided by Safegraph and can now be obtained from Veraset, a company that aggregates anonymized location data from a suite of smartphone applications. The smartphone data records ``pings" denoting where a specific smartphone is located at a particular point in time. Pings are logged at irregular time intervals, whenever a participating smartphone application requests location information. The modal time between consecutive pings associated with a device is roughly 10 minutes. Our smartphone data covers more than 50 million smartphones, spanning the continental US, in a 10-month period from February 2017 to November 2017. While the dataset contains geolocation information from only a subset of all smartphones, previous studies have found it highly representative of the United States on numerous demographic dimensions (\citealt{long2020political}).

We link the smartphone data to two other data sources: 1) police station location data published by the Department of Homeland Security, verified with the city's open data portal and google maps data, and 2) building rooftop geofence data provided by Microsoft, enabling us to associate each police station's latitude-longitude location to a geofence that delineates the convex hull of a building's rooftop boundary. To identify patrol officers in local city neighborhoods, we include police stations categorized as patrol stations, as headquarters, or as unspecified police facilities, resulting in a total of 330 stations across 21 of America’s largest cities. A description of other data sources and data cleaning process can be found in Appendix A1. It is important to note the selection of the cities in our sample was based on jurisdictional population and the physical construction of police buildings.  Our sample was not determined by the investment the department chose to make in electronic monitoring of officers, or a department's decision to release the data publicly or enter into a research agreements with external parties (see \cite{Goeletal2017} for a discussion of these issues in the context of measuring police bias). 

\subsection{Measuring Police Presence}

We infer whether a smartphone belongs to an officer by linking smartphone data to police stations' geofences in several steps.
First, if a specific smartphone is observed in a police station geofence at least five days in a month, we identify it as belonging to a police employee in that month. We next infer each smartphone user’s “home” as the smartphone's modal Geohash-7 (a 152m $\times$ 152m grid) during a five-month period when the device is not at any police stations. We identify two home locations separately for the early and the latter half of the year to account for a potential summer move. Then, we identify patrol officers by looking for a specific pattern: leaving home, traveling to a police station, moving around the city (without returning home), returning to the police station, and then going home. The movements of that smartphone between the first and the last station visits are assumed to be the actual locations of a patrol officer while working a ``shift." We require that shifts are bracketed by home visits that are no more than 24 hours apart, and are no shorter than four hours.\footnote{All results in our analysis are highly robust to limiting our sample to 8 to 12 hour shifts, requiring shifts bracketed by home visits no longer than 18 hours and excluding shifts with long hours spent within the police stations.  These results are available on request.} Under this definition, our officer smartphone GPS data sample consists of 10,131 officers that have at least one shift, with the mean of shift lengths being 8.08 hours.\footnote{Figure \ref{fig:mode} in the Online Appendix displays the spatial pattern of pings for one likely LAPD officer.}

To measure police presence in all census block groups (``neighborhoods") within the city's jurisdiction, we look at officers’ smartphone pings outside of the police stations when officers are ``on shift" in the month where the device has at least 5 days' presence. We conceptualize police presence in a city neighborhood as the number of officer-hours spent in the neighborhood. Specifically, we match police officers’ ping locations to block groups, exclude pings moving faster than 50 mph, and assign the duration of each ping as half of the time between its previous and next ping.\footnote{Using other constructs of police presence yields qualitatively and quantitatively similar results.  Replications of our analysis using the number of distinct officer shifts, alternate (or no) speed thresholds, and alternate lengths of shifts or time between home visits are available on request.} We then compute the sum of officer-hours from all officers' pings observed in the block group as police hours across the ten-month period. Where police spend time on patrol is highly non-uniform, and as our later regression analysis will confirm, is strongly correlated with demographics in ways that produce large racial disparities.

\subsection{Validity Check}

Our study focuses on America’s largest cities. While our data do not capture the universe of police officers in a city, our estimates of the number of officers in a city satisfy many tests of face validity as a measure of police presence. The number of officer devices that we observe across US cities is highly correlated with FBI estimates of police force size ($\rho$ = 0.98 for total count measures, $\rho$ = 0.64 for per capita measures).\footnote{Appendix Figure \ref{fig:officers_count} plots the specific values for each city.}  Further, we can probabilistically impute each device's ``race” using its home census block's racial composition to estimate each police department's racial composition.  There is essentially a one for one unconditional relationship between the imputed race of the police departments in our sample and the racial composition reported by the department in the 2016 Law Enforcement Management and Administration Statistics (LEMAS); conditional on the racial composition of the city, a one percentage point increase in our estimate of the percent of the police force that is White (Black, Hispanic, Asian) is associated with a 0.6 (0.5, 0.9, 0.6) percentage point increase in the reported percent of the force that is White (Black, Hispanic, Asian) in the LEMAS. \footnote {Appendix Figure \ref{fig:LEMAS_compare} plots the raw data. The p-value testing whether the slope between the smartphone GPS measures and LEMAS measure of racial composition is equal to 1 is 0.946 for Black, 0.890 for White, and 0.099 for Hispanic. The slopes between the two estimates for the share of Asian is significantly different from 1, though Asians account for only 2.5\% of the police force across the cities in LEMAS. Table \ref{tab:LEMAS} in the Appendix further shows that this significant correlation is not simply driven by cross-city variation in racial composition.}

We conduct an additional residence-based validity check in New York City, in which public records provide summary data on where NYPD officers live at the zip code level. We compare the NYPD's official records on the number of officers who live in a zip code with the number of officers that our smartphone GPS measures estimate ``live" in that zip code. There is a strong and positive correlation ($\rho$ = 0.71) between official NYPD records and our smartphone-based measures.\footnote{Figure \ref{fig:NYPD_counts} in the Appendix plots the zip code level data.} 

\begin{figure}[H]
    \centering
    \caption{Correlation Between \% Black and Officers per capita in a City}
    \label{fig:officer_black}
    \includegraphics[width=\linewidth]{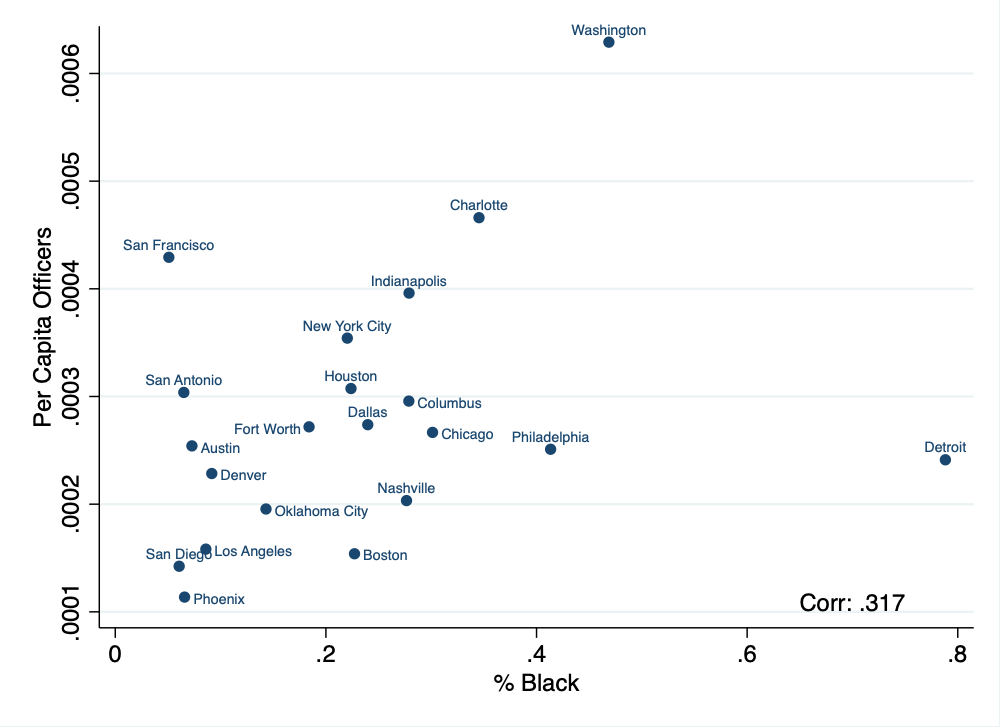}
    {\footnotesize 
    \textit{Notes:} \textit{Per capita officers} is defined as the number of likely patrol officers on “shift” divided by the city population. We identify patrol officers on “shift” by looking for a specific pattern in smartphones that visit at least 5 days in a month: Leaving “home”, traveling to a police station, moving around the city (without returning home), returning to the police station, and then going home. The correlation coefficient between the two measures is reported.
    }
\end{figure}

There is a well-established positive correlation between the fraction of a city population that is Black and the number of sworn police officers per capita (\citealt{CarmichaelKent2014,stults2007racial}). A basic test of construct validity is therefore to test if we observe a similar pattern in our data. Figure \ref{fig:officer_black} plots per capita patrol officers (i.e. smartphones that have at least one shift) against the share of Black population in the 21 cities, replicating the positive correlation between the fraction of city residents who are Black and our measure of total officers per capita. Our GPS-based measure of police presence has significant predictive power on down-funnel measures of police actions, such as stops and arrests. After adjusting for nonlinearity, the correlation between our measure of police presence and the number of arrests—which we observe in six cities—ranges from 0.46 (Washington) to 0.68 (Austin). Similar positive and significant correlations for police stops for nine cities with publicly available geocoded records on police stops are observed. \footnote{Appendix Figures \ref{fig:arrests} and \ref{fig:stops} plot these city by city graphs.}

\section{Results}

\subsection{Neighborhood Correlates of Police Presence }

Understanding how police provide services to people from different racial groups is important from both an equity and an efficiency perspective, and our data are uniquely suited to provide new evidence on this issue. Within each neighborhood, we measure the concentration of different racial/ethnic groups as the ratio of the percent of neighborhood residents who report being in a particular category divided by the percent of city residents who report the same.  This reflects the observation that exactly what constitutes a “more” Asian, Black, or Hispanic neighborhood is likely different in Detroit than in San Francisco. One convenient feature of our relative share measure is that one unit increase relative to the mean roughly corresponds to a neighborhood with twice as many Asian, Black, or Hispanic residents as in the city as a whole. Table \ref{tab:sumstat} in the Appendix shows summary statistics for police presence measures as well as neighborhood correlates.\footnote{Results measuring racial share in an absolute way, or using a series of dummies to indicate that the majority of residents identify themselves as a particular group are qualitatively identical, and available on request.} 

Table \ref{tab:main} presents our estimates of the spatial determinates of policing in America’s largest cities. Our smartphone GPS data reveal a strong relationship between the racial and ethnic composition of a neighborhood and police presence. Relative to the average city neighborhood, our coefficients in Table \ref{tab:main} suggest police spend 8.2\% more time in a place where the fraction of residents who are Black is twice the city's share, 5.2\% more time in a place with twice the share of Hispanic residents, and 1.4\% more time in a place where twice as many neighborhood residents are Asian as in their respective city as a whole.\footnote{The elasticity for a arsinh-transformed $y$ is $\sqrt{\frac{1+\bar{y}^2}{\bar{y}}}\bar{x}\beta$, hence for Black concentration, the calculation of the elasticity is 1.0007*1.023*0.0801 = 8.2\%, for Hispanic concentration, the elasticity is 1.0007*0.944*0.0554  5.2\%, and for Asian concentration, the elasticity is 1.0007*0.895*0.0155 = 1.4\%.
} 

Why do these disparities exist? Differences in where police spend their time can reflect decisions made by individual officers and department-level directives, both of which involve assessing the ``need” for police presence in an area. Such department policies, officer decisions, and residential demand for police presence can all be related to the racial composition of the neighborhood. We use a multivariate regression framework to provide insight into why police may tend to spend more time in places with relatively more Asian, Black, and Hispanic residents.  

We begin by introducing proxies for residential demand.  If officers spend more time in places where there are more people, variation in population density that is correlated with race may contribute to spatial differences in policing. Location decisions can also indicate residential demand for police presence. Residents may request that officers respond to crimes, and in particularly disadvantaged neighborhoods, police officers may be one of the few remaining providers of any social service (\citealt{Lum2021}). Racial disparities in police presence may therefore stem from racial inequity in the quality of non-policing institutions. 

We draw on existing social science literature to approximate components of civilian demand for police.  A lack of educational opportunity and well-paying jobs are established root causes of crime (\citealt{messner1997political}). Of course, neighborhoods where residents have low incomes but high social capital (e.g. high degrees of social cohesion and community engagement) are places where police rarely need to respond to acts of violence or property destruction (\citealt{sampson1999systematic}). Following \cite{martin2015measuring}, we measure social capital using the fraction of 2010 census forms returned by residents. Finally, police officers go where violent crime exists. We estimate the crime-driven demand for police based on the location of homicides known to the police. While imperfect and sparse, police records of homicides are generally thought to be the most accurate, in the sense that reporting of homicides is unlikely to be as influenced by police presence as reporting of other types of crime (\citealt{Levitt1998}). We calculate the distance from the neighborhood center to the closest homicide in 2016, treating these rare events as an extreme expression of underlying social issues, implicitly assuming that this distance is negatively correlated with exposure to other types of crime. Additionally, we control for the number of homicides in 2016, by neighborhood, to account for potential high-end variation in crime. 

Alternate measures of demand for policing, specifically using additional years of homicide data and 311 calls for service in New York City (\citealt{shah_laforest2021}) are explored in the Online Appendix - none lead to substantively different conclusions. We also compare how police presence varies with residential composition when we exclude times of day, and places, where the modal person on the street may not live there.  In the Online Appendix, we show that our findings are qualitatively identical when we model police presence during non-working hours (excluding 9 am - 5 pm), and in New York City when we exclude census block groups in tourist destinations.  Finally, we show that the relationship between exposure to police presence and the relative composition of the block group that is Black or Hispanic is highest during the middle of an officer's shift.

In column 2 of Table \ref{tab:main}, we condition our estimates of local police presence in different types of U.S. neighborhoods on measures of density, socioeconomics, social cohesion, and violence. Differential residential demand for police presence, some of which is created by decisions made in other policy domains, explains approximately 26\% of the disparate exposure of people living in relatively Black neighborhoods, 42\% of the disparate exposure of people living in relatively Hispanic neighborhoods, and can explain all of the additional exposure of people living in relatively Asian neighborhoods—even suggesting that more Asian neighborhoods have less police presence than one might expect based on social conditions. The residual correlation between racial composition and police presence in column 2 reflects decisions at the city, police command, and officer level.

In column 3, we include city-specific fixed effects, differencing out any preference of officials in cities with more concentrated non-White populations for a particular type of policing that may contribute to observed disparities in policing. Focusing on variation in neighborhoods within cities reduces the estimated extra time officers spend in more relatively Black neighborhoods by an additional 60\%, and reduces the lighter policing of more Asian neighborhoods by 11\%. Across large cities, officers spend more time in neighborhoods with relatively more Hispanic people, but we do not observe this pattern within cities. 

Diversifying the officer ranks is one city-level policy that is central to many police reform efforts. With this in mind, we compare how disparities in police presence vary with the racial composition of a city's police force. We do this in two ways: including the mean-centered interaction between the relative share of Black residents and the share of police officers that are Black in column 4, and interactions with both the share of police supervisors and patrol officers that are Black in column 5.\footnote{Appendix Figure \ref{fig:supervisor} reveals substantial variation in the Black share in police officers and supervisors relative to the city's Black population across cities, and a meaningful difference between the share of Black officers and supervisors, despite a high correlation between the two measures.} Both departmental measures are divided by the share of the city that is Black. Column 4 suggests the additional exposure to police in Black neighborhoods is smaller in cities with a larger share of Black officers. Column 5 implies that conditional on the composition of patrol officers, when the share of Black command staff is higher, there is less additional police presence in Black neighborhoods. In contrast to previous studies (e.g. \citealt{Ba_al2021b}), conditional on the share of command that is Black, we find that jurisdictions that employ more Black officers are also jurisdictions with more exposure of people in Black neighborhoods to police, both across and within cities (Column 6). While correlational in nature, our findings suggest that efforts to hire more Black police officers, without parallel efforts to retain and promote those officers, may not reduce disparities in how the public is policed.

\subsection{Cross-city variation in correlates of police presence}

\begin{figure}[H]
    \centering
    \includegraphics[width = \linewidth]{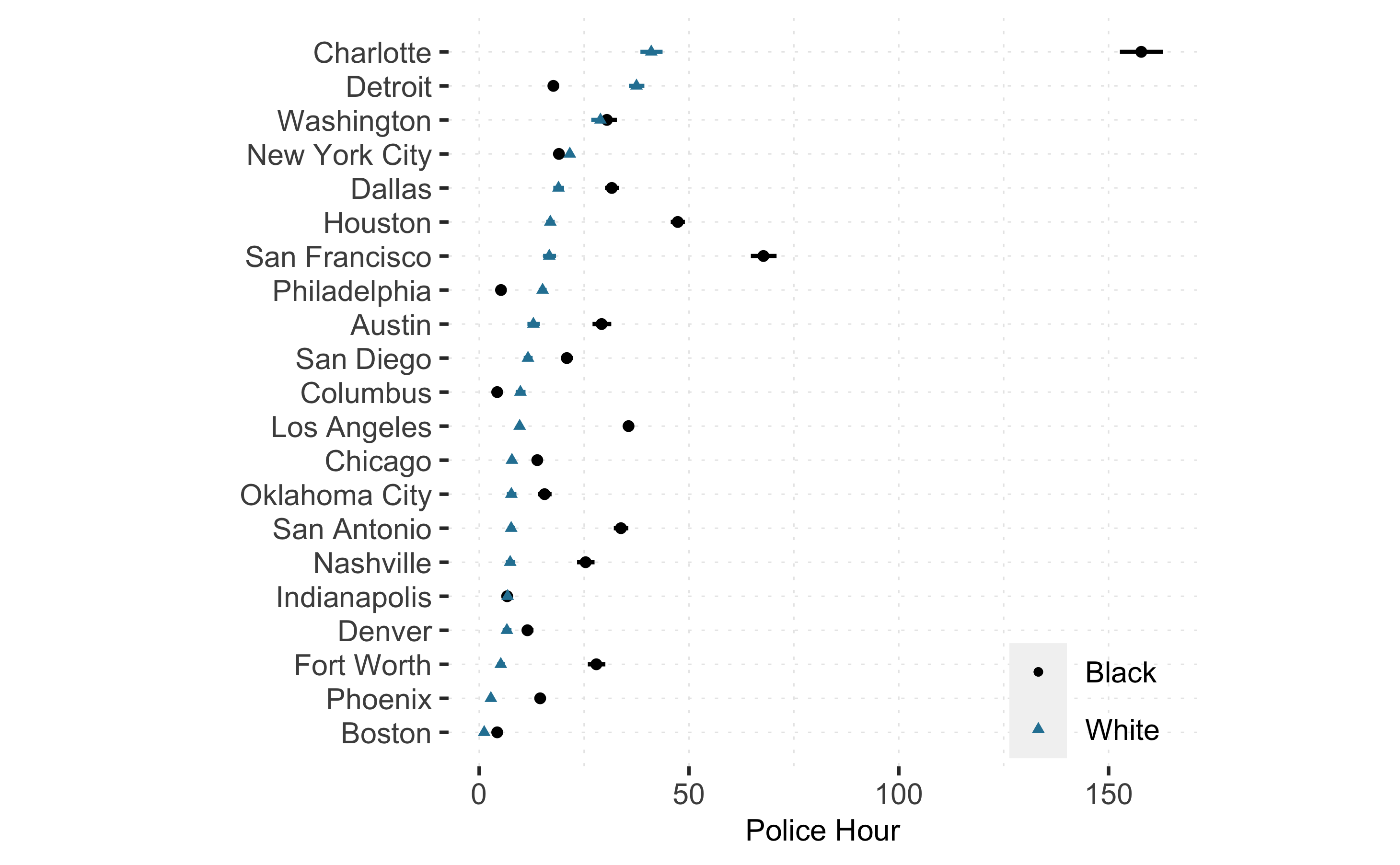}
    \caption{Police Exposure in Blackest and Whitest Neighborhoods}
    \label{fig:hour}
        {\footnotesize
    \textit{Notes:}
 This figure plots the average police hours observed in the Blackest (Whitest) neighborhoods in a city, defined as the block groups where share of Black (White) residents is over the 95th percentile of the city's distribution.
}
\end{figure}

Our findings suggest substantial differences in the level of ambient police presence in non-White neighborhoods across the United States, and this difference is largest in Black neighborhoods. Given this, and the long and fraught history of the policing of Black people in the United States, in this section we report the difference in 2017 police presence in neighborhoods with the highest concentrations of Black residents and highest concentrations of White residents. First, in Figure \ref{fig:hour}, we show police presence in neighborhoods with the greatest concentration of Black and White residents, respectively, to highlight the range of disparities in Black-White neighborhood policing across major US cities.  There is little difference in the amount of time that police spend in the ``most White" and ``most Black" neighborhoods in Boston, but over 100 more hours of total policing in the ``most Black" neighborhoods in Charlotte than in the ``most White."  

\begin{figure}[H]
    \centering
    \includegraphics[width=\linewidth]{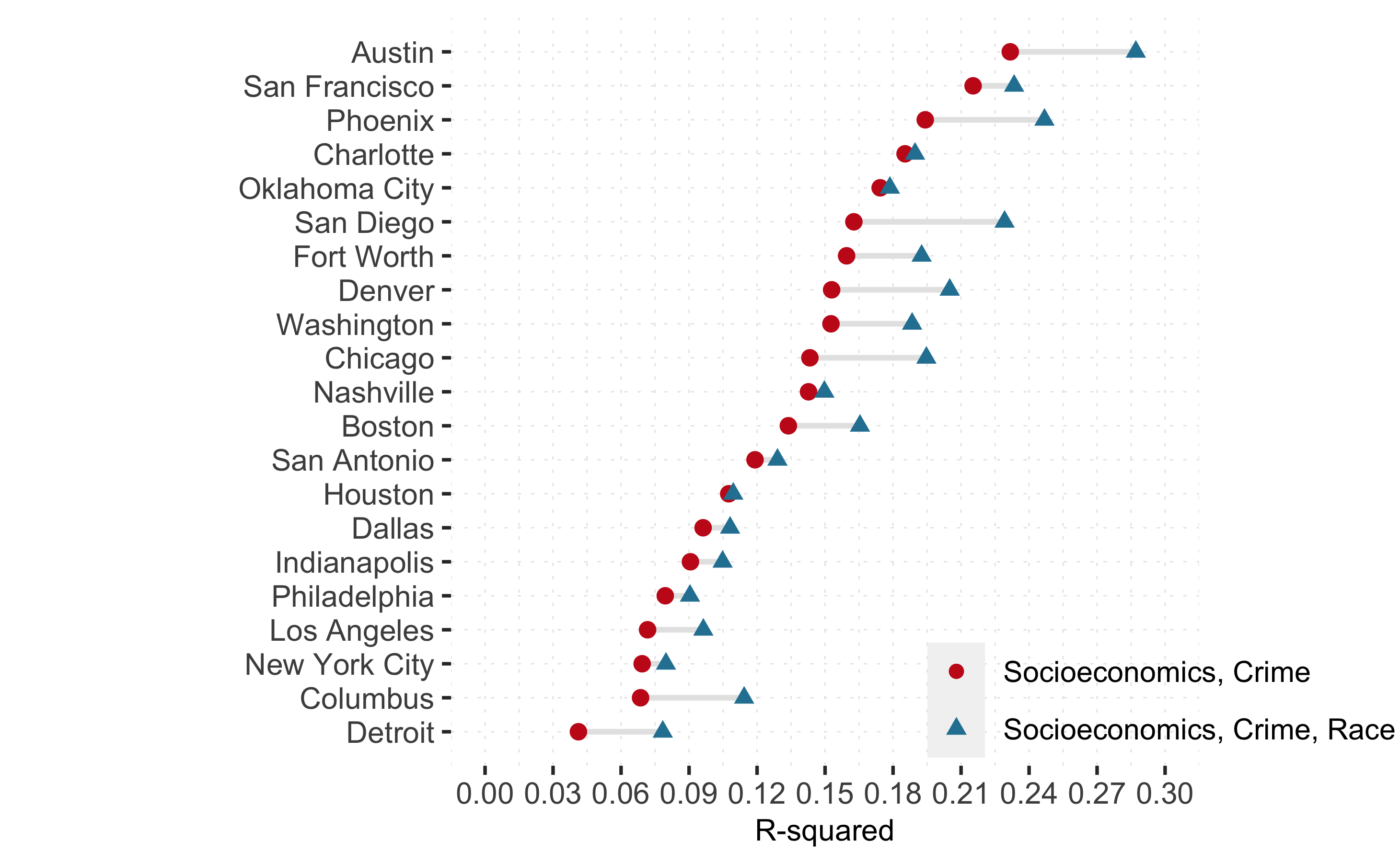}
    \caption{Variance of Police Hours Explained by Socioeconomics, Crime, and Race}
    \label{fig:cities}
    {\footnotesize
    \textit{Notes:} Socioeconomic variables include Log Population, \% College Graduates, Median Household Income, Census Form Return Rate, Crime variables include Distance to Nearest Homicide and Homicide Count in 2016. Race variables include relative Black share, relative Hispanic share, and relative Asian share, defined as the ratio of percent Black (Hispanic, Asian) in the block group to percent Black (Hispanic, Asian) in the city.
}
\end{figure}

Of course, these disparities can have many sources.
In Figure \ref{fig:cities}, we plot, for each city, how much of the spatial variation in police presence can be explained by spatial variation in our proxies for ``demand" for police, and how much the explanatory power of our models increases when we add controls for relative racial composition.  This shows the extent to which Black-White disparities in exposure to police can persist even when considering spatial differences in socioeconomic status - which in some cases may reflect historical and contemporary race-based social and economic inequality. We document substantial variation across cities in the role of this structural inequality in explaining policing disparities. For example, while Figure \ref{fig:hour} reveals large differences in the ambient police exposure of Black and White Charlotte residents, Figure \ref{fig:cities} reveals that spatial disparities in socioeconomic status explain almost all of these differences. These structural disparities in Charlotte are city-level issues that cannot be addressed solely by the city's police department. In contrast, racial disparities in police presence are absolutely smaller in Austin, but incorporating Black, Hispanic, and Asian residential patterns increases the amount of spatial variation in police presence that we can explain in that city by 27\%.  This suggests substantially more scope for changes in police policy to reduce criminal justice disparities in Austin, TX.

\subsection{Using Police Presence to Understand Police Enforcement} 

The empirical observation that police are more ambiently present in Non-White neighborhoods provides support for the construct validity of our data, as this correlation has been repeatedly demonstrated at the city level (\citealt{CarmichaelKent2014}). When taken in the context of existing qualitative and legal scholarship on modern policing, this also raises equity concerns. 

Past research highlights how the increased use of surveillance (either remotely or in-person) by law enforcement and algorithmic policing can reinforce existing racial disparities, even in the absence of differential criminal behavior by residents (\citealt{Ba_al2021b, brayne2020predict}). Our measure of police presence can quantify the extent to which variation in ambient police presence, rather than differences in the enforcement behavior of officers in different neighborhood contexts, can predict racial disparities in police enforcement. 

Consistent with studies of down-funnel measures of policing, Table \ref{tab:arrest} confirms that in the six cities (New York City, Los Angeles, Chicago, Dallas, Austin, Washington) for which we have both police presence and arrest data, officers spend more time, and make more arrests in places with higher relative Black shares. Note that the Black-White disparity in police presence is 27\% smaller, the Hispanic-White disparity 20\% smaller, and the Asian-White disparity 165\% larger, in this sub-sample of cities that make geocoded arrest data public. In addition, officers appear to make almost 24\% more arrests per hour present in neighborhoods where the share of residents who are Black is twice the city's share.\footnote{In contrast to our findings regarding the source of Black-White arrest disparities, smartphone data do not suggest that differences in police presence are an important source of Hispanic-White arrest disparities.}  This could be due to Becker-style discrimination, where police use different standards to determine if people in different groups are suspicious enough to warrant an arrest, or to differences in where police officers spend time in these neighborhoods.\footnote{This type of Becker-style discrimination is not necessarily illegal; Illinois v Wardlow, 528 U. S. 119 (2000) established that officers can use the predetermined designation of an area as ``high crime" in determining how likely it is someone has (or is) engaged in crime, creating a legal basis for a stop.  If places with more Black or Hispanic residents are more likely to be known to police as ``high crime" places, then this would lower the standard of individualized suspicion needed to make a constitutionally permissible stop.} Whatever the source, this disparity in the propensity of an officer to make an arrest explains less than half of the disparity in the total number of arrests made.  This implies that the added time that police spend in Black neighborhoods may be a central source of Black-White disparities in arrests, in addition to an officer's decision in a particular encounter. Online Appendix Table \ref{tab:stop} also reveals a highly similar pattern regarding stop disparities. Consistent with \cite{Meares2015}, our results suggest that, to reduce disparities in criminal justice, reducing the scope for racial bias both in specific civilian encounters and in police directives detailing where officers go and who they surveil may be warranted. 

\section{Conclusion}

We conclude by noting that a positive correlation in the provision of policing and the concentration of Black residents contrasts with documented spatial patterns of other institutional investment in neighborhoods with concentrated Black populations. Census tracts where more of the residential population is Black are less, not more, likely to have a large grocery store, nearby hospital, or local banking services (\citealt{walker2010disparities}, \citealt{deyoung2008commercial}, \citealt{lieberman2020disparities}, \citealt{yearby2018racial}). During the 2016 election, \cite{chen2019racial} found that voting lines moved more slowly in places with larger Black populations, suggesting under-investment in polling services in places where we observe larger investments in ambient policing.

Our data are well suited to further research on policing in the United States. First, smartphone location data provide insight into officer presence in communities that traditional measures of policing cannot fully capture. Measuring officer presence informs estimates of which communities are at risk of more serious police encounters, like arrest or the use of lethal force. Second, our smartphone location data do not depend on software purchased by or developed for a particular policing agency, allowing us to map officer locations in cities across the United States using a consistent methodology. This is an advantage over technologies like AVL and body cameras, because it provides enhanced visibility into the unreported and highly discretionary activities of police officers at work. Finally, data on where officers actually spend their time grants researchers and practitioners new abilities to understand patterns in police presence and track the implementation of departmental policies that shape the provision of public safety. Thereby, our data may serve as a baseline measure to objectively evaluate the intensity of down-funnel police actions like stops and arrests within neighborhoods across major U.S. cities.


\newpage

\bibliographystyle{econ}
\bibliography{REFERENCES}

\newpage

\section*{Figures and Tables}

\begin{table}[H]
\centering
\caption{Disparities in Neighborhood Police Exposure}
\label{tab:main}
\resizebox{\linewidth}{!}{
\begin{tabular}{lcccccc} \hline
 & (1) & (2) & (3) & (4) & (5) & (6) \\
   VARIABLES & \multicolumn{6}{c}{Police Exposure in a Census Block Group: arsinh(Hours)} \\ \midrule
Relative Black Share & 0.0801*** & 0.0596*** & 0.0235** & 0.0543*** & 0.0713*** & 0.0398*** \\
 & (0.00731) & (0.00814) & (0.00794) & (0.00857) & (0.00904) & (0.00895) \\
Police: Relative Black &  &  &  & 0.361*** & -0.734*** &  \\
 &  &  &  & (0.0338) & (0.0795) &  \\
Relative Black Share X Police: Relative Black &  &  &  & -0.0446+ & 0.158** & 0.160** \\
 &  &  &  & (0.0230) & (0.0563) & (0.0530) \\
Supervisor: Relative Black &  &  &  &  & 0.648*** &  \\
 &  &  &  &  & (0.0407) &  \\
Relative Black Share X Supervisor: Relative Black &  &  &  &  & -0.118*** & -0.102*** \\
 &  &  &  &  & (0.0282) & (0.0267) \\
Relative Hispanic Share & 0.0554*** & 0.0320** & -0.00977 & 0.0153 & 0.0311* & -0.000532 \\
 & (0.00999) & (0.0116) & (0.0104) & (0.0118) & (0.0121) & (0.0112) \\
Relative Asian Share & 0.0155* & -0.0239*** & -0.0213*** & -0.0264*** & -0.0168* & -0.0159* \\
 & (0.00748) & (0.00607) & (0.00584) & (0.00635) & (0.00678) & (0.00645) \\
Log Population &  & 0.540*** & 0.426*** & 0.548*** & 0.563*** & 0.455*** \\
 &  & (0.0211) & (0.0212) & (0.0219) & (0.0223) & (0.0226) \\
\% College Graduates &  & 1.083*** & 0.918*** & 1.036*** & 1.226*** & 1.042*** \\
 &  & (0.0615) & (0.0602) & (0.0629) & (0.0640) & (0.0633) \\
Median Household Income (1K) &  & -0.00425*** & -0.00452*** & -0.00510*** & -0.00469*** & -0.00424*** \\
 &  & (0.000394) & (0.000395) & (0.000405) & (0.000407) & (0.000408) \\
Census Form Return Rate &  & -0.635*** & -1.343*** & -0.446*** & -0.723*** & -1.400*** \\
 &  & (0.125) & (0.127) & (0.131) & (0.134) & (0.137) \\
Distance to nearest 2016 homicide (km) &  & -0.0983*** & -0.125*** & -0.0718*** & -0.0808*** & -0.114*** \\
 &  & (0.00624) & (0.00660) & (0.00657) & (0.00718) & (0.00755) \\
Homicide Count 2016 &  & 0.200*** & 0.205*** & 0.214*** & 0.224*** & 0.207*** \\
 &  & (0.0206) & (0.0200) & (0.0212) & (0.0215) & (0.0212) \\
 &  &  &  &  &  &  \\
Observations & 23,682 & 22,521 & 22,521 & 20,961 & 20,112 & 20,112 \\
R-squared & 0.005 & 0.069 & 0.166 & 0.073 & 0.090 & 0.157 \\
 Fixed effects & NA & NA & City & NA & NA & City \\ \hline
   \end{tabular}}

   \footnotesize{\textit{Notes:} This table presents OLS estimates of exposure disparities among census block groups (BGs) across (Column 1,2,4,5) and within 21 large cities (Column 3, 6). All race variables are mean-centered, relative Black (Hispanic, Asian) shares are defined as the ratio of \% Black (Hispanic, Asian) in a BG to the \% in that city, Police (Supervisor) Relative Black defined as the ratio of \% Black of a department's sworn officers (supervisors) to the \% in that city. The dependent variable is police hours observed in a BG (excluding pings moving faster than 50 mph), transformed into arsinh values. Household income is measured in thousands of dollars, census return rates range from 0-1. Robust standard errors are reported in parentheses, and standard errors clustered at the tract level are reported in the SI Appendix. Results are qualitatively and quantitatively similar to running all regressions with log dependent variable and dropping zero-valued observations, or clustering at the tract level, and are available on request. *** p$<$0.001, ** p$<$0.01, * p$<$0.05, + p$<$0.1}
   
\end{table}

\newpage

\begin{table}[H]
\centering
\caption{Disparities in Neighborhood Police Exposure and Downstream Disparities}
\label{tab:arrest}
\resizebox{\linewidth}{!}{
\begin{tabular}{lcccccc} \hline
 & (1) & (2) & (3) & (4) & (5) & (6) \\
VARIABLES & arsinh(Hours) & arsinh(Arrests) & arsinh(Arrests/Hour) & arsinh(Hours) & arsinh(Arrests) & arsinh(Arrests/Hour) \\ \midrule
Relative Black Share & 0.0584*** & 0.325*** & 0.221*** & 0.0955*** & 0.180*** & 0.0759*** \\
 & (0.00926) & (0.0105) & (0.00957) & (0.0110) & (0.0109) & (0.0108) \\
Relative Hispanic Share & 0.0448** & 0.433*** & 0.299*** & 0.122*** & 0.217*** & 0.0701*** \\
 & (0.0150) & (0.0136) & (0.0135) & (0.0200) & (0.0178) & (0.0181) \\
Relative Asian Share & 0.0411*** & 0.0278** & -0.0176* & 0.0153 & -0.0265** & -0.0368*** \\
 & (0.00964) & (0.00979) & (0.00773) & (0.00981) & (0.00956) & (0.00810) \\
Log Population &  &  &  & 0.593*** & 0.554*** & -0.0886*** \\
 &  &  &  & (0.0313) & (0.0273) & (0.0234) \\
\% College Graduates &  &  &  & 1.343*** & 0.261*** & -0.924*** \\
 &  &  &  & (0.0839) & (0.0772) & (0.0675) \\
Median Household Income (1K) &  &  &  & -0.00354*** & -0.00326*** & 0.000489 \\
 &  &  &  & (0.000503) & (0.000467) & (0.000395) \\
Census Form Return Rate &  &  &  & -0.285+ & -1.935*** & -1.310*** \\
 &  &  &  & (0.166) & (0.148) & (0.150) \\
Distance to nearest 2016 homicide (km) &  &  &  & -0.0502*** & -0.230*** & -0.130*** \\
 &  &  &  & (0.0127) & (0.0122) & (0.0114) \\
Homicide Count 2016 &  &  &  & 0.261*** & 0.333*** & 0.0602* \\
 &  &  &  & (0.0280) & (0.0222) & (0.0237) \\
 &  &  &  &  &  &  \\
Observations & 12,748 & 12,748 & 12,705 & 12,098 & 12,098 & 12,059 \\
 R-squared & 0.003 & 0.137 & 0.087 & 0.074 & 0.253 & 0.133 \\ \bottomrule
\end{tabular}}
    \footnotesize{\textit{Notes:} The sample of six cities with arrest data include: New York City, Los Angeles, Chicago, Dallas, Austin, Washington.  All race variables are mean-centered, relative Black (Hispanic, Asian) shares are defined as the ratio of \% Black (Hispanic, Asian) in a BG to the \% in that city, Police (Supervisor) Relative Black defined as the ratio of \% Black of a department's sworn officers (supervisors) to the \% in that city. The dependent variables are: police hours observed in a BGs (excluding pings moving faster than 50 mph, mean 26.9), number of arrests in that BG (mean 40.1), and the ratio of those two measures (mean 10.2), all transformed into arsinh values. Household income is measured in thousands of dollars, census return rates range from 0-1. Robust standard errors are reported in parentheses. Results are qualitatively and quantitatively similar to running all regressions with log independent values and dropping zero-valued observations, or clustering at the tract level, and are available on request. *** p$<$0.001, ** p$<$0.01, * p$<$0.05, + p$<$0.1}
\end{table}

\end{document}


\begin{titlepage}
 \title{\textbf{Online Appendix} \\
 Smartphone Data Reveal Neighborhood-Level Racial Disparities in Police Presence} 
 \author{M. Keith Chen, Katherine L. Christensen, Elicia John, \\ Emily Owens, Yilin Zhuo\thanks{Chen: UCLA Anderson School of Management, Los Angeles, CA 90095, USA, keith.chen@naderson.ucla.edu. 
 Christensen: Indiana University, Kelley School of Business, Bloomington, IN 47405, USA, kachris@iu.edu. John: American University, Kogod School of Business, Washington, DC 20016, USA, eliciaj@american.edu. Owens: Department of Criminology, Law \& Society, Department of Economics, University of California Irvine, Irvine, CA 92697, USA, egowens@uci.edu. Zhuo: UCLA Anderson School of Management, Los Angeles, CA 90095, USA, yilin.zhuo.phd@anderson.ucla.edu. Acknowledgements: We thank Milind Rajavasireddy for excellent research assistance, as well as Bocar Ba, Kitt Carpenter, Amanda Geller, Michael Norton, Broderick Turner, and seminar participants in the Virtual Crime Economics (ViCE) Seminar and the USC-UMich-UVA Online Seminar on Law and Economics for helpful comments.}}
 \date{\today}
 \maketitle
 \setcounter{page}{0}
 \thispagestyle{empty}
 \end{titlepage}
 \pagebreak \newpage
 
\doublespacing

\renewcommand{\thesection}{\Alph{section}}

\section{Other Data Sources}

\textbf{Census demographics data.} Census block group, and city characteristics data come from American Community Survey (ACS) 2013-2017 5-year estimates. We collect data on each block group’s racial composition (\% Black, Hispanic, and Asian), population, median household income, percent college graduates, and census form mail return rate. We also collect city level data on racial composition (\% Black, White, Hispanic, and Asian).

\textbf{Homicide data.} Homicide data is collected by The Washington Post and covers homicide information (including latitude-longitude location, arrest decision, victim demographics) in 50 of the largest U.S. cities from 2007 to 2017 (\citealt{wapo}). For several cities, the homicide data is not available for the whole decade: for example, in New York City, data are provided in 2016 and 2017 only; for San Antonio, data are only available between 2013 to 2016. The definition of homicide follows the FBI’s Uniform Crime Reporting Program, including murder and nonnegligent manslaughter while excluding suicides, accidents, justifiable homicides, and deaths caused by negligence. We use records of homicides to measure crime-driven demand for policing given the high accuracy of homicide reporting. 

\textbf{The Law Enforcement Management and Administrative Statistics (LEMAS) data.} The LEMAS data contains information on police officers’ demographics, salaries, and functions, as well as on agencies’ duties, structures, and policies for 3499 local law enforcement agencies in 2016 (\citealt{lemas}). We obtain the racial composition of full time sworn officers and supervisors for 21 cities’ police departments to compare with the imputed race of smartphone users. Among the 21 cities, the Indianapolis Metropolitan police department is not included in the LEMAS data, while the Phoenix Police Department and the San Antonio Police Department have missing data on officers’ and supervisors’ race respectively.

\textbf{FBI Uniform Crime Report (UCR) - Law Enforcement Officers Killed or Assaulted (LEOKA) data.} UCR-LEOKA data contains measures of officers that are killed or assaulted and total officer employment as of October 1st of each year at the departmental level (\citealt{leoka}). We compare the police employee counts (with and without arrest powers) in the 2017 UCR-LEOKA data with our counts of qualified smartphones that visit the police stations at least five times a month.

\textbf{NYPD Officer Home Zip Code data.} Data on NYPD police officers’ home zip code comes from \cite{bell21} through submission of a Freedom of Information Law (FOIL) request to the NYPD. The data reports the number of police officers that live in a specific zip code and patrol in a specific precinct. We calculate the total number of police officers that live in a zip code across all precincts to compare with the police officer counts that we infer to ``live'' in that zipcode from the smartphone location data.

\textbf{Police Action data.} We collect 6 cities' geocoded data on police arrests in 2017 from each city's open data portal.\footnote{The 6 cities are: New York City, Los Angeles, Chicago, Dallas, Austin, Washington.} We collect geocoded data on police stops in nine cities from multiple sources, including open data portals for New York City, Philadelphia and Denver, Stanford Open Policing Project (\citealt{pierson2020large}) for Columbus, Nashville, Houston, San Antonio, and Oklahoma City and \cite{Ba_al2021b} for Chicago. We collect 2017 stop data for most cities, and for cities in which 2017 data are not available, we use data closest to 2017: for Chicago, we use data in 2015; for Columbus and Oklahoma City, we use data in 2016. We match the latitude-longitude location of a police action to a census block group and aggregate the total number of stops or arrests during a year in a block group. Note that a small fraction of police action data are missing location information. While the missing records usually account for less than 5\% of the observations for most cities, 13.49\% of the stop records have missing location information for the Chicago Police Department.

\section{Alternative Crime-driven Demand Measures}



In this section, we explore the robustness of policing disparity estimates to alternative crime-driven demand measures. In the main paper, we measure crime-driven demand using homicide data in 2016 due to concerns on relevance and data constraints (homicide data earlier than 2016 are not available for New York City, which has the most block groups of any city in our sample). In Table \ref{tab:noNYC13-16}, we exclude New York City and use distance to the nearest homicide and average homicide counts between 2013 and 2016- this explains all disparity in police hours in Black neighborhoods in the remaining 20 cities. In Table \ref{tab:nyc13-16}, we add New York City back to the sample and fill in the missing values with the 2016 measures for New York City, and find estimates that are more similar to our main results. The differences in estimates between Table \ref{tab:noNYC13-16} and Table \ref{tab:nyc13-16} imply a larger racial disparity in police presence in New York City. In addition, it is worth noting that while including information on older homicides does allow researchers to differentiate between neighborhoods without homicides in 2016, it is not obvious that police ``should" do the same. Given the potential negative consequences of police interaction, particularly for young Black people \citep{Rios2011}, failing to update patrol patterns to reflect current, rather than past, violence may itself be a component of anti-Black bias in addition to a proxy for neighborhood demand for police. 

To provide a direct measure of demand for police services as well as suspicion of criminal activity, in Table \ref{tab:311calls}, we control for the number of 311 calls in New York City where the geocoded 311 calls data are made publicly available. In the case of New York City, we do not find evidence suggesting that the number of 311 calls explain the police presence disparity in Black neighborhoods, regardless of controlling for the total number of calls (Column 3), or calls handled specifically by NYPD (Column 4), or calls handled by the nine major agencies (Column 5). In contrast, conditional on neighborhood socioeconomic characteristics, the number of 311 calls explains away 30\% of the enhanced police time in Hispanic neighborhoods, and all additional police time in Asian neighborhoods.

\section{Sensitivity to Visitors' Foot Traffic}

In this section, we demonstrate that our results are not sensitive to foot traffic from non-residents in two ways. First, we examine police presence during non-working hours by excluding pings between 9 am to 5 pm on weekdays, shown in Table \ref{tab:nonwork}. We observe a strikingly similar pattern as in Table 1, suggesting that the estimates are not driven by daytime foot traffic. Second, we complement the above analysis by removing block groups that are likely to have large levels of visitor foot traffic in one city, New York City, that accounts for the largest number of block groups (N = 6,226) among the 21 cities. We exclude block groups in Precinct 1 (Wall Street), 6 (the West Village), 8 (Penn Station, Grand Central), 14 (Midtown South) and 18 (Midtown North). Comparing the estimates of exposure disparities where we include every block group in NYPD precincts (column 1-2) or exclude block groups in five NYPD precincts (column 3-4) in Table \ref{tab:NYC}, suggests that our results are insensitive to the exclusion of precincts with potentially high levels of non-residential foot traffic.
 
\section{Disparities over the course of a shift}

\begin{figure}[H]
 \centering
 \caption{Racial Disparity in Police Presence by Hour of the Shift}
 \label{fig:shift_hr_blk_disparity}
 \begin{subfigure}{0.7\linewidth}
 \centering
 \caption{Black-White Disparity}
 \includegraphics[width=\linewidth]{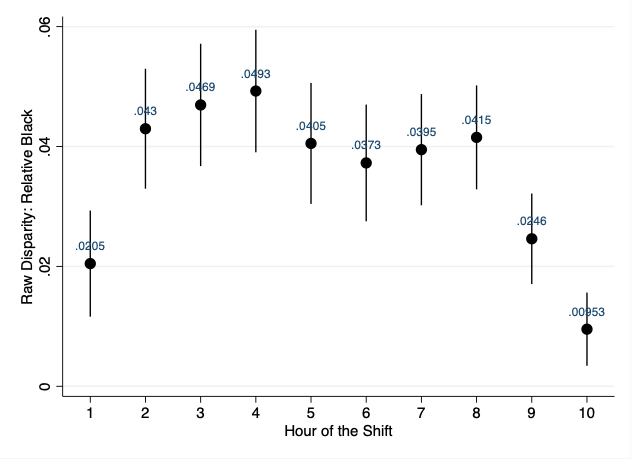}
 \end{subfigure}
 \hfill
 \begin{subfigure}{0.7\linewidth}
 \centering
 \caption{Hispanic-White Disparity}
 \includegraphics[width=\linewidth]{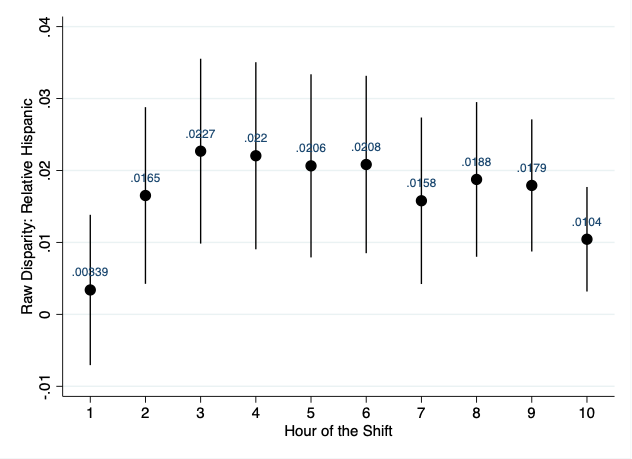}
 \end{subfigure}
 
 {\footnotesize 
 \textit{Notes:} Figure plots coefficients of relative Black (Hispanic) share from a regression where police presence in each hour of the shift is regressed against the relative share of Black, Hispanic and Asian population. 
 }
\end{figure}

Officers begin each shift at a station and, after receiving specific instructions about their daily tasks (in a process known as ``roll call"), leave to patrol their beat with relatively little real-time oversight. Enforcement activity generally peaks midway through an officer's shift, suggesting that the way officers spend their patrol time may vary over the course of a day (\citealt{chalfinGoncalves2021}). Appendix Figure \ref{fig:shift_hr_blk_disparity} plot how the share of time officers spend in more Hispanic and more Black places increases as their shift rolls out. In the first hour of officers' shifts, the racial and ethnic composition of a neighborhood is not associated with where they patrol. The difference between how much time officers spend in more Hispanic versus Whiter places increases from the first hour of the shift through the third hour. In places where more Black people live, the disparities in police time are most pronounced halfway through a shift and then decline. 

\newpage

\section*{Figures and Tables}

\begin{figure}[H]
 \caption{Spatial Pattern of Pings of a Smartphone Observed in LAPD}
 \label{fig:mode}
 \centering
 \includegraphics[width=\linewidth]{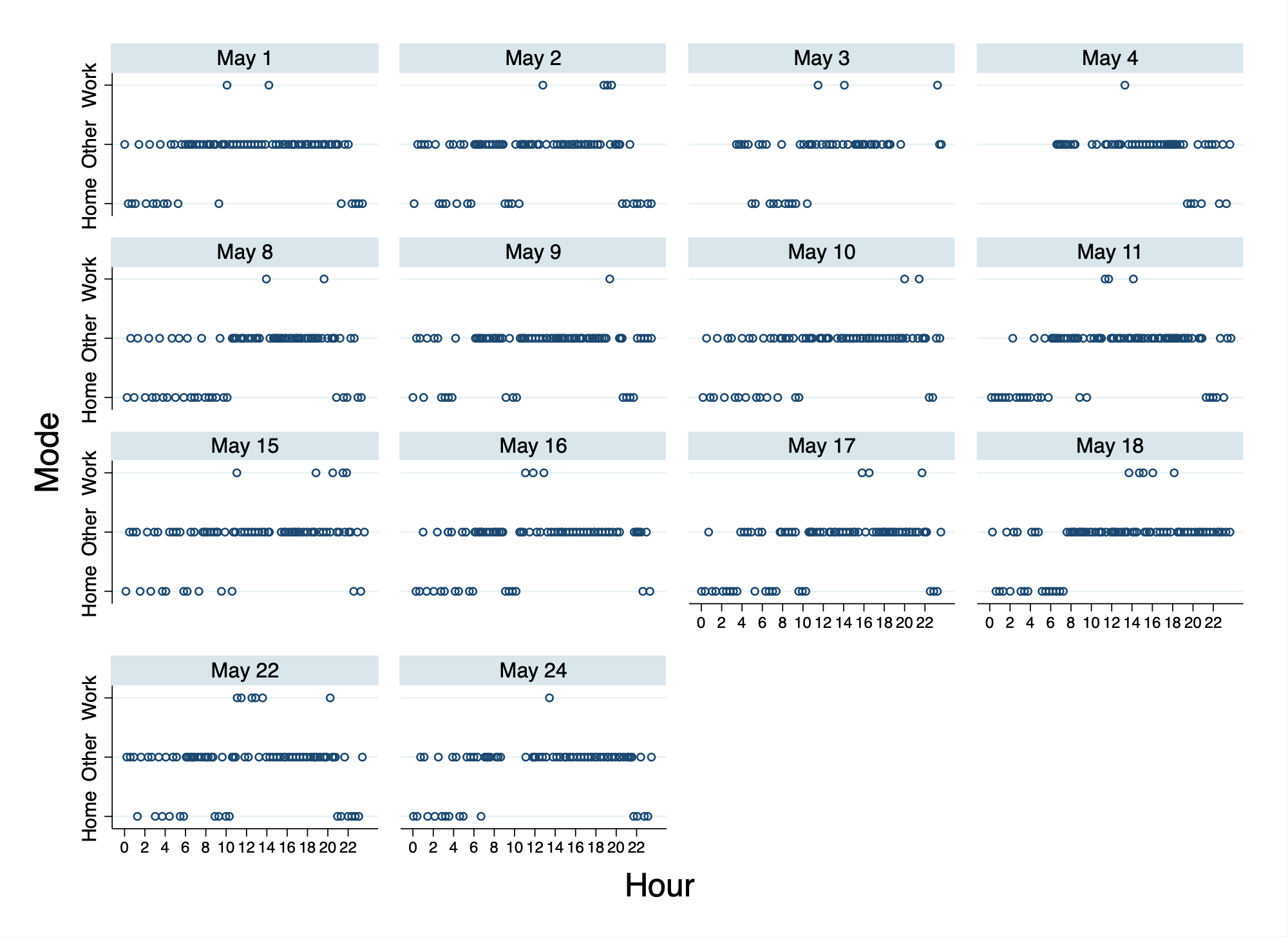}
 
 {\footnotesize
 \textit{Notes:} The spatial pattern of smartphone pings is categorized as either Home, Other, or Work. Smartphone is “at home” if the ping location is at the Home Geohash-7 (a 152 x 152 m grid); “at Work” if the ping location is in any police stations’ building boundaries. Pings observed at locations other than “Home” and “Work” are classified as “Other”.
 }

\end{figure}

\begin{figure}[H]
 \centering
 \caption{Police Officer Validation at the City Level}
 \label{fig:officers_count}
 \begin{subfigure}[h]{0.7\linewidth}
 \centering
 \caption{UCR Employee Counts and Smartphone Counts}
 \includegraphics[width=\linewidth]{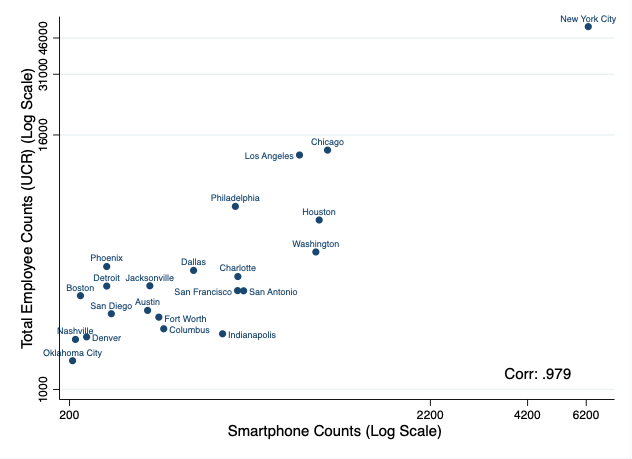}
 \end{subfigure}
 \hfill
 \begin{subfigure}[h]{0.7\linewidth}
 \centering
 \caption{UCR Employee and Smartphone (Per Capita Value)}
 \includegraphics[width=\linewidth]{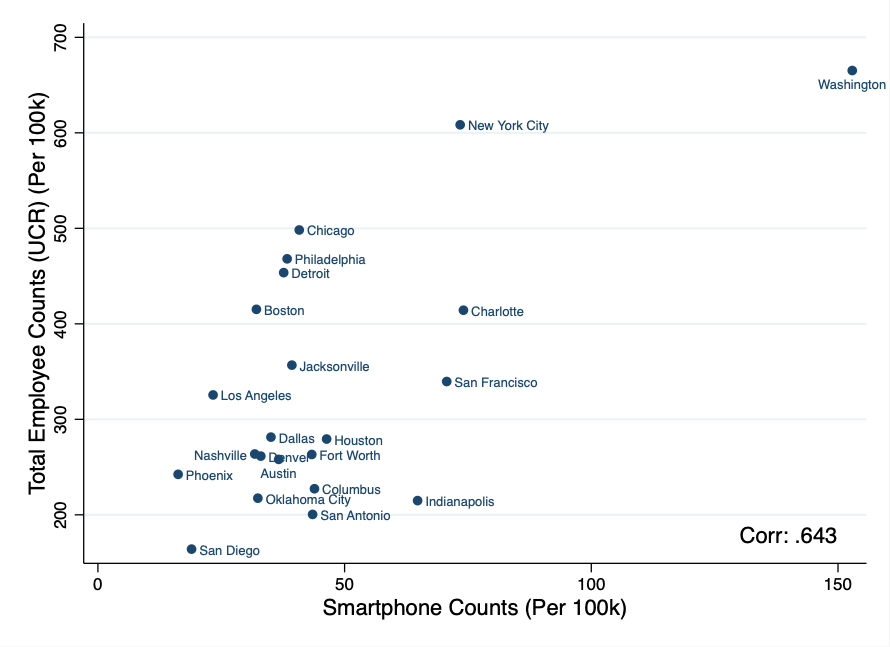}
 \end{subfigure}
 
 {\footnotesize
 \textit{Notes:} \textit{Total Employee Counts} on the y-axis reports the number of officers (with or without arrest power) in each city’s police department on October 1st, 2017. \textit{Smartphones Counts} reports the number of smartphones visiting the city’s local police stations for at least 5 days in any one month during Feb 2017 to Nov 2017. Correlation coefficient between the two measures is reported.
 }
\end{figure}

\begin{figure}[H]
 \centering
 \caption{LEMAS Police Force Racial Composition vs. Smartphone Racial Composition}
 \label{fig:LEMAS_compare}
 \includegraphics[width=\linewidth]{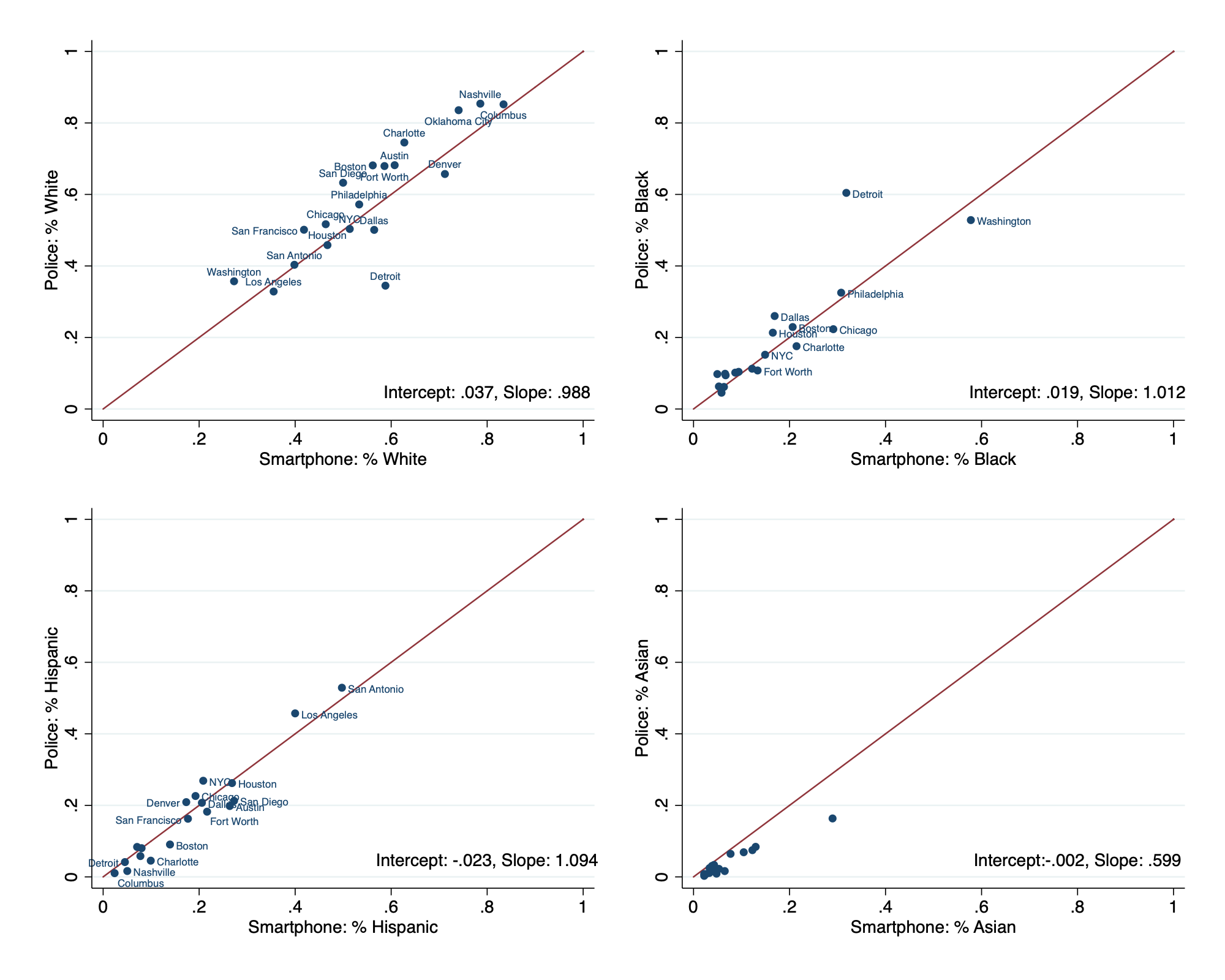}
 
 {\footnotesize 
 \textit{Notes: } Police \% White (Black, Hispanic, Asian) represents measures of racial composition from LEMAS data. Smartphone: \% White (Black, Hispanic, Asian) denotes the smartphone-imputed racial composition based on home blocks. 
} 
\end{figure}

\begin{figure}[H]
 \centering
 \caption{Police officer validation: residence-based check for NYPD officers at the zip code level.}
 \label{fig:NYPD_counts}
 \includegraphics[width=0.7\linewidth]{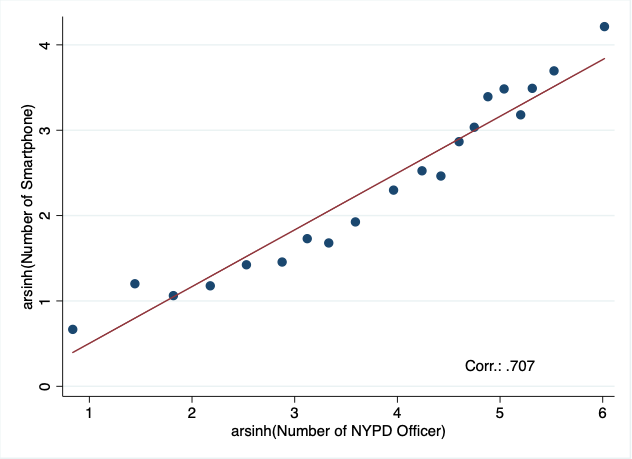}
 
 {\footnotesize
 \textit{Notes:} This figure presents a binned scatter plot of the number of smartphones from NYPD that we infer "live" in a zip code vs. the actual number of NYPD police officers living in a zip code, both transformed in arsinh values. We include all zip codes in the FOIL request data, with zip codes grouped into 20 equal-sized bins. Correlation coefficient between the two measures (in arsinh values) is reported.
}
\end{figure}

\newpage

\begin{figure}[H]
 \centering
 \caption{The Number of Arrests vs. Police Hours Across Block Groups}
 \label{fig:arrests}
 \includegraphics[width=\linewidth]{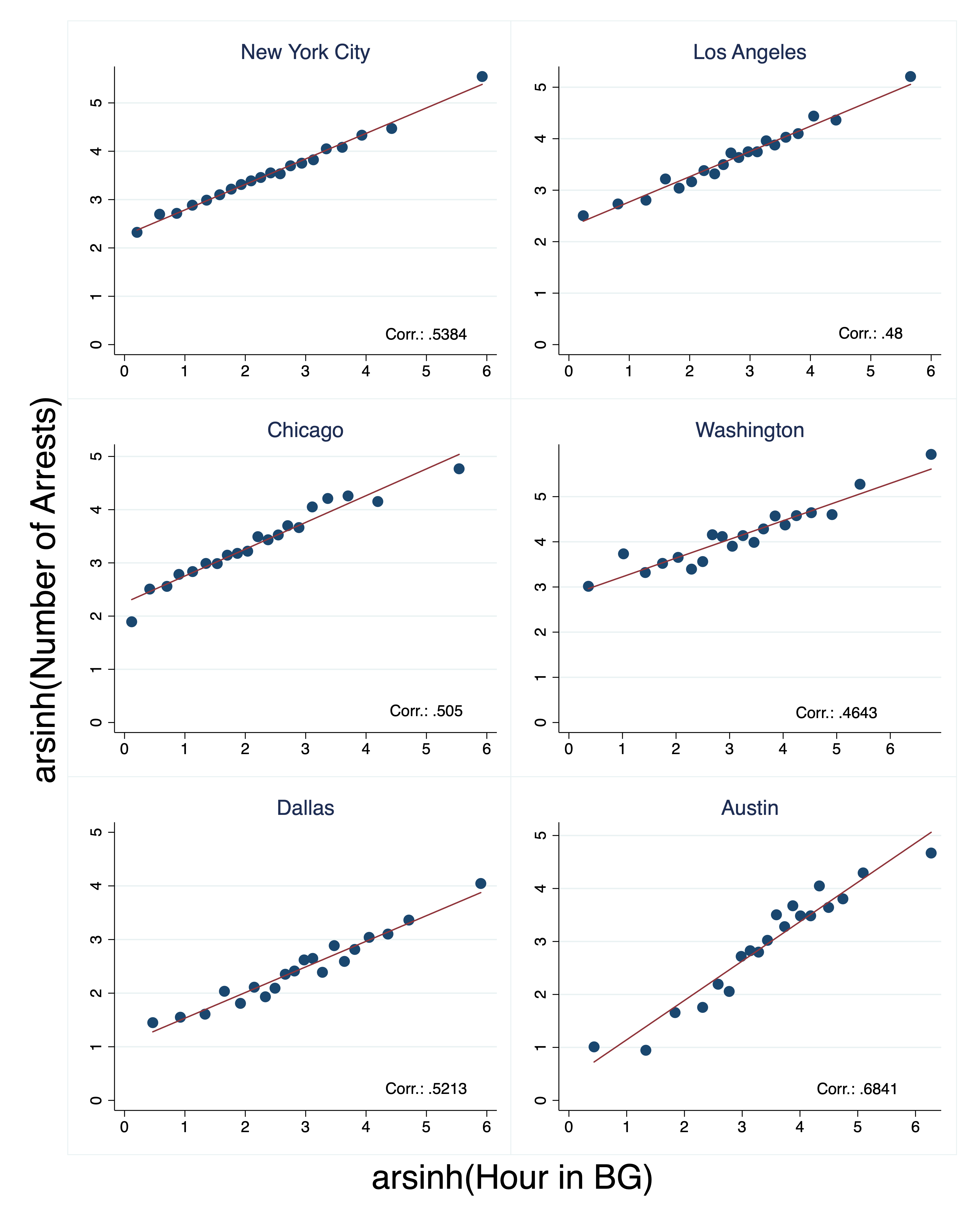}
 {\footnotesize \textit{Notes:} Each panel presents a binned scatter plot of the number of arrests vs. the police hours observed in the block groups, with both variables measured in arsinh values. Block groups are grouped into 20 equal size bins. Correlation coefficient between the two measures (in arsinh values) is reported in each panel. 
 }
\end{figure}

\begin{figure}[H]
 \centering
 \caption{The Number of Stops vs. Police Hours Across Block Groups}
 \label{fig:stops}
 \includegraphics[width=\linewidth]{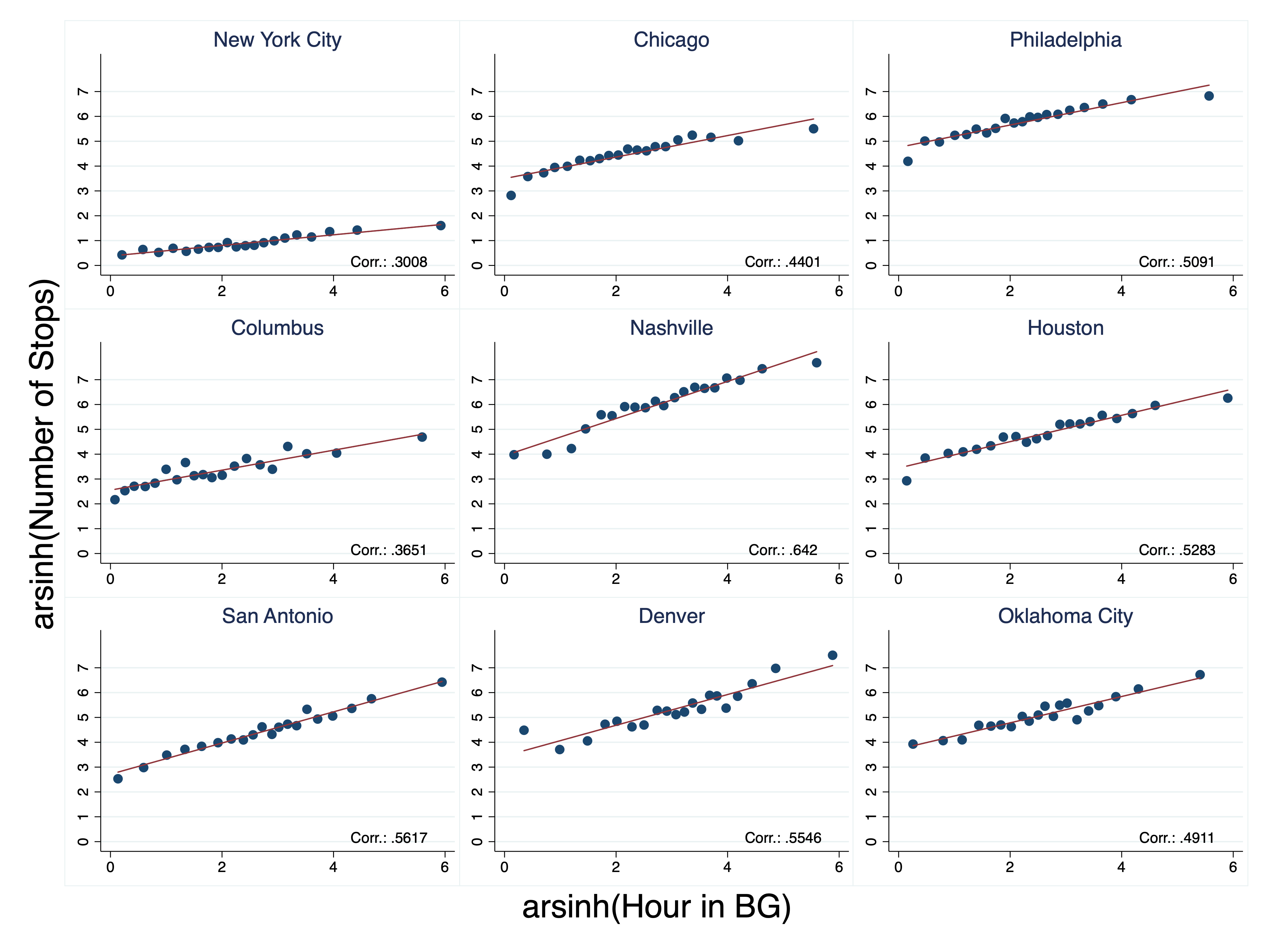}
 {\footnotesize 
 \textit{Notes:} Each panel presents a binned scatter plot of number of stops vs. the police hours observed in the block groups, with both variables transformed in arsinh values. Block groups are grouped into 20 equal-sized bins. Correlation coefficient between the two measures (in arsinh values) is reported in each panel. 
 }
\end{figure}

\begin{figure}[H]
 \centering
 \caption{Supervisor Relative Black Share vs. Officer Relative Black Share}
 \label{fig:supervisor}
 \includegraphics[width=0.7\linewidth]{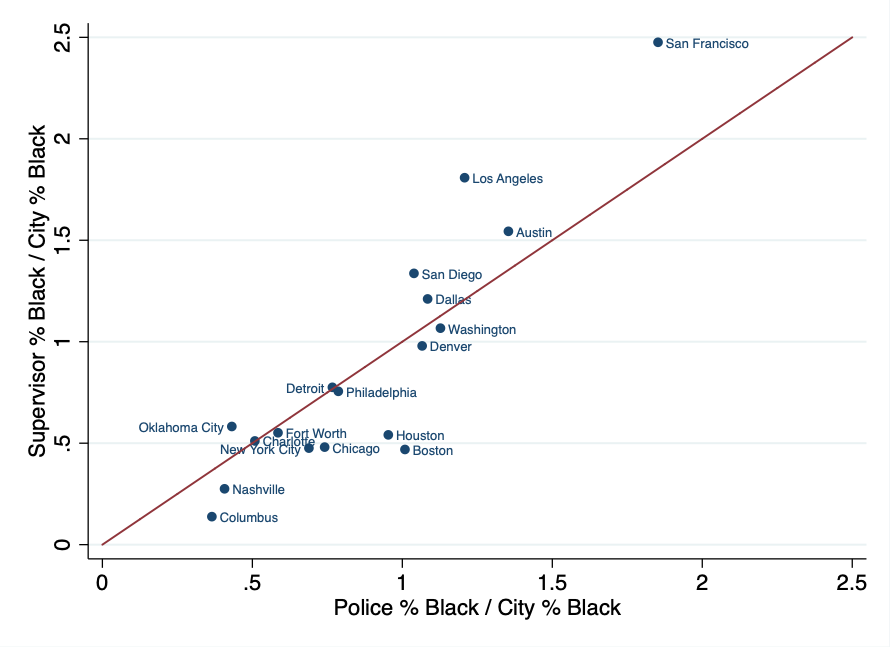}

\end{figure}

\begin{table}[H]
 \centering
 \caption{Racial Composition: Smartphone Measure vs. LEMAS}
 \label{tab:LEMAS}
 \resizebox{0.75\linewidth}{!}{%
 \begin{tabular}{lcccc} \hline
 & (1) & (2) & (3) & (4) \\
 VARIABLES & Police: \% White & Police: \% Black & Police: \% Hispanic & Police: \% Asian \\ \hline
 & & & & \\
Smartphone: \% White & 0.583*** & & & \\
 & (0.0960) & & & \\
City \% White & 0.757*** & & & \\
 & (0.116) & & & \\
Smartphone: \% Black & & 0.525*** & & \\
 & & (0.130) & & \\
City \% Black & & 0.472*** & & \\
 & & (0.143) & & \\
Smartphone: \% Hispanic & & & 0.931*** & \\
 & & & (0.181) & \\
City \% Hispanic & & & 0.137 & \\
 & & & (0.125) & \\
Smartphone: \% Asian & & & & 0.637*** \\
 & & & & (0.168) \\
City \% Asian & & & & -0.0325 \\
 & & & & (0.146) \\
Constant & -0.0311 & -0.0117 & -0.0298** & -0.00230 \\
 & (0.0478) & (0.0180) & (0.0112) & (0.00331) \\
 & & & & \\
Observations & 19 & 19 & 19 & 19 \\
 R-squared & 0.909 & 0.899 & 0.936 & 0.933 \\ \hline
 \end{tabular}}
 
 {\footnotesize 
 \textit{Notes: } Police \% White (Black, Hispanic, Asian) represents measures of racial composition from LEMAS data. Smartphone: \% White (Black, Hispanic, Asian) denotes the smartphone-imputed racial composition based on home blocks. City \% White (Black, Hispanic, Asian) denotes the share of population that is identified as White (Black, Hispanic, Asian) in the city. Robust standard errors are reported in parentheses: *** p$<$0.001, ** p$<$0.01, * p$<$0.05, + p$<$0.1.

} 
 \end{table}
 
\begin{table}[H]
\centering 
\caption{Summary Statistics \label{tab:sumstat}}
\resizebox{\linewidth}{!}{%
\begin{tabular}{l c c c c c }\hline\hline
\multicolumn{1}{c}{Variable} & Obs & Mean & Std. Dev.
 & Min & Max \\ \hline
 \textbf{Police Presence:} & & & & & \\ 
Hour & 23799 & 26.685 & 201.754 & 0 & 14683.32 \\
arsinh(Hour) & 23799 & 2.483 & 1.428 & 0 & 10.288 \\
Number of Shifts & 23799 & 70.34 & 129.557 & 0 & 5306 \\
arsinh(Number of Shifts) & 23799 & 4.246 & 1.261 & 0 & 9.27 \\
 \textbf{Neighborhood Characteristics:} & & & & & \\ 
\% Black & 23682 & .237 & .31 & 0 & 1 \\
\% Hispanic & 23682 & .287 & .284 & 0 & 1 \\
\% Asian & 23682 & .084 & .137 & 0 & .983 \\
Relative Black Share & 23682 & 1.023 & 1.284 & 0 & 12.696 \\
Relative Hispanic Share & 23682 & .944 & .981 & 0 & 12.891 \\
Relative Asian Share & 23682 & .895 & 1.537 & 0 & 64.464 \\
Black Majority & 23682 & .199 & .399 & 0 & 1 \\
White Majority & 23682 & .361 & .48 & 0 & 1 \\
Hispanic Majority & 23682 & .235 & .424 & 0 & 1 \\
Asian Majority & 23682 & .027 & .162 & 0 & 1 \\
Population & 23799 & 1425.74 & 820.84 & 0 & 18369 \\
\% College Graduates & 23679 & .338 & .251 & 0 & 1 \\
Median Household Income (1K) & 22526 & 62.553 & 38.174 & 2.499 & 250.001 \\
Census Form Return Rate & 23671 & .736 & .088 & 0 & 1 \\
Distance to nearest 2016 homicide (km) & 23799 & 1.331 & 1.612 & .001 & 23.759  \\
Homicide Count 2016 & 23799 & .152 & .472 & 0 & 7  \\
\hline\end{tabular}}
\end{table}

\newpage
\begin{table}[H]
\centering
\caption{Disparities in Neighborhood Police Exposure (Measuring Homicide from 2013-2016, Excluding NYC)}
\label{tab:noNYC13-16}
\resizebox{\linewidth}{!}{%
\begin{tabular}{lccccc} \hline
 & (1) & (2) & (3) & (4) & (5) \\
 VARIABLES & \multicolumn{5}{c}{Police Exposure in a Census Block Group: arsinh(Hours)} \\ \hline
 & & & & & \\
 Relative Black Share & 0.0897*** & 0.0455*** & -0.00799 & 0.0401*** & 0.0697*** \\
 & (0.00853) & (0.00934) & (0.00901) & (0.0107) & (0.0121) \\
Police: Relative Black & & & & 0.385*** & -0.670*** \\
 & & & & (0.0382) & (0.0803) \\
Relative Black Share X Police: Relative Black & & & & -0.0536* & 0.141* \\
 & & & & (0.0259) & (0.0563) \\
Supervisor: Relative Black & & & & & 0.637*** \\
 & & & & & (0.0411) \\
Relative Black Share X Supervisor: Relative Black & & & & & -0.124*** \\
 & & & & & (0.0296) \\
Relative Hispanic Share & 0.0584*** & 0.0101 & -0.0532*** & -0.00958 & 0.00681 \\
 & (0.0115) & (0.0128) & (0.0109) & (0.0130) & (0.0135) \\
Relative Asian Share & 0.0218* & -0.0242*** & -0.0163* & -0.0252*** & -0.0163* \\
 & (0.00897) & (0.00662) & (0.00636) & (0.00698) & (0.00752) \\
Log Population & & 0.539*** & 0.386*** & 0.554*** & 0.577*** \\
 & & (0.0232) & (0.0232) & (0.0243) & (0.0249) \\
\% College Graduates & & 1.026*** & 0.738*** & 0.967*** & 1.206*** \\
 & & (0.0715) & (0.0688) & (0.0733) & (0.0752) \\
Median Household Income (1K) & & -0.00456*** & -0.00526*** & -0.00608*** & -0.00552*** \\
 & & (0.000476) & (0.000460) & (0.000499) & (0.000504) \\
Census Form Return Rate & & -1.152*** & -1.827*** & -0.753*** & -1.010*** \\
 & & (0.161) & (0.154) & (0.168) & (0.174) \\
Avg 13-16 Homicide Count & & 0.237*** & 0.261*** & 0.262*** & 0.280*** \\
 & & (0.0195) & (0.0187) & (0.0205) & (0.0211) \\
Distance to nearest 13-16 homicide (km) & & -0.0917*** & -0.144*** & -0.0362** & -0.0599*** \\
 & & (0.0108) & (0.0106) & (0.0126) & (0.0130) \\
Constant & 2.496*** & -0.587** & 1.188*** & -0.884*** & -0.991*** \\
 & (0.0109) & (0.201) & (0.202) & (0.210) & (0.214) \\
 & & & & & \\
Observations & 17,456 & 16,700 & 16,700 & 15,140 & 14,291 \\
R-squared & 0.007 & 0.076 & 0.211 & 0.085 & 0.107 \\
 Fixed effects & NA & NA & City & NA & NA \\ \hline
 \end{tabular}
}
 {\footnotesize 
 
 \textit{Notes: } This table presents OLS estimates of exposure disparity among census block groups (BGs) across 20 cities (Column 1,2,4,5) and within cities (Column 3). All race variables are mean-centered, relative Black (Hispanic, Asian) shares defined as ratio of \% Black (Hispanic, Asian) in BG to \% in city, Police (Supervisor) Relative Black defined as ratio of \% Black of sworn officers (supervisors) to \% in city. The dependent variable is police hours observed in BGs (excluding pings moving faster than 25 mph), transformed in arsinh values. Household income is measured in thousands of dollars, census return rates range from 0-1. Robust standard errors are reported in parentheses: *** p$<$0.001, ** p$<$0.01, * p$<$0.05, + p$<$0.1.
}

\end{table}

\newpage
\begin{table}[H]
\centering
\caption{Disparities in Neighborhood Police Exposure (Measuring Homicide from 2013-2016, Including NYC)}
\label{tab:nyc13-16}
\resizebox{\linewidth}{!}{%
\begin{tabular}{lccccc} \hline
 & (1) & (2) & (3) & (4) & (5) \\
 VARIABLES & \multicolumn{5}{c}{Police Exposure in a Census Block Group: arsinh(Hours)} \\ \hline
 & & & & & \\
Relative Black Share & 0.0801*** & 0.0552*** & 0.0155* & 0.0510*** & 0.0685*** \\
 & (0.00731) & (0.00810) & (0.00791) & (0.00852) & (0.00899) \\
Police: Relative Black & & & & 0.314*** & -0.773*** \\
 & & & & (0.0340) & (0.0793) \\
Relative Black Share X Police: Relative Black & & & & -0.0426+ & 0.183** \\
 & & & & (0.0227) & (0.0557) \\
Supervisor: Relative Black & & & & & 0.638*** \\
 & & & & & (0.0404) \\
Relative Black Share X Supervisor: Relative Black & & & & & -0.130*** \\
 & & & & & (0.0282) \\
Relative Hispanic Share & 0.0554*** & 0.0403*** & -0.00755 & 0.0242* & 0.0392** \\
 & (0.00999) & (0.0119) & (0.0104) & (0.0121) & (0.0124) \\
Relative Asian Share & 0.0155* & -0.0219*** & -0.0197*** & -0.0238*** & -0.0135* \\
 & (0.00748) & (0.00607) & (0.00582) & (0.00635) & (0.00685) \\
Log Population & & 0.520*** & 0.406*** & 0.534*** & 0.553*** \\
 & & (0.0210) & (0.0210) & (0.0217) & (0.0222) \\
\% College Graduates & & 1.161*** & 0.979*** & 1.118*** & 1.307*** \\
 & & (0.0619) & (0.0603) & (0.0632) & (0.0644) \\
Median Household Income (1K) & & -0.00393*** & -0.00442*** & -0.00471*** & -0.00422*** \\
 & & (0.000396) & (0.000394) & (0.000409) & (0.000411) \\
Census Form Return Rate & & -0.821*** & -1.275*** & -0.594*** & -0.869*** \\
 & & (0.125) & (0.128) & (0.130) & (0.134) \\
Avg 13-16 Homicide Count & & 0.270*** & 0.296*** & 0.288*** & 0.289*** \\
 & & (0.0179) & (0.0180) & (0.0185) & (0.0188) \\
Distance to nearest 13-16 homicide (km) & & -0.112*** & -0.166*** & -0.0739*** & -0.0951*** \\
 & & (0.00979) & (0.00980) & (0.0108) & (0.0113) \\
 & & & & & \\
Observations & 23,682 & 22,521 & 22,521 & 20,961 & 20,112 \\
R-squared & 0.005 & 0.072 & 0.171 & 0.078 & 0.095 \\
 Fixed effects & NA & NA & City & NA & NA \\ \hline
 \end{tabular}
}
 {\footnotesize 
 
 \textit{Notes: } This table presents OLS estimates of exposure disparity among census block groups (BGs) across 21 cities (Column 1,2,4,5) and within cities (Column 3). All race variables are mean-centered, relative Black (Hispanic, Asian) shares defined as ratio of \% Black (Hispanic, Asian) in BG to \% in city, Police (Supervisor) Relative Black defined as ratio of \% Black of sworn officers (supervisors) to \% in city. The dependent variable is police hours observed in BGs (excluding pings moving faster than 25 mph), transformed in arsinh values. Household income is measured in thousands of dollars, census return rates range from 0-1. Robust standard errors are reported in parentheses: *** p$<$0.001, ** p$<$0.01, * p$<$0.05, + p$<$0.1.
}

\end{table}

\newpage
\begin{table}[H]
\centering
\caption{Disparities in Neighborhood Police Exposure (Control for Number of 311 Calls)}
\label{tab:311calls}
\resizebox{\linewidth}{!}{%
\begin{tabular}{lccccc} \hline
 & (1) & (2) & (3) & (4) & (5) \\
 VARIABLES & \multicolumn{5}{c}{Police Exposure in a Census Block Group: arsinh(Hours)} \\ \hline
 & & & & & \\
Relative Black Share & 0.0427** & 0.127*** & 0.126*** & 0.133*** & 0.145*** \\
 & (0.0147) & (0.0185) & (0.0176) & (0.0182) & (0.0173) \\
Relative Hispanic Share & 0.0375+ & 0.212*** & 0.0772* & 0.114*** & 0.0861** \\
 & (0.0207) & (0.0309) & (0.0307) & (0.0316) & (0.0301) \\
Relative Asian Share & -0.0198 & 0.0374* & 0.0506** & 0.0317+ & 0.0107 \\
 & (0.0154) & (0.0169) & (0.0158) & (0.0164) & (0.0143) \\
Log Population & & 0.455*** & 0.113* & 0.242*** & 0.0894* \\
 & & (0.0488) & (0.0459) & (0.0486) & (0.0403) \\
\% College Graduates & & 1.707*** & 1.158*** & 1.464*** & 0.185 \\
 & & (0.136) & (0.128) & (0.133) & (0.132) \\
Median Household Income (1K) & & -0.00150* & -0.000998 & -0.000891 & -0.00239*** \\
 & & (0.000735) & (0.000680) & (0.000706) & (0.000665) \\
Census Form Return Rate & & -0.322 & 1.276*** & 0.871** & 0.817** \\
 & & (0.250) & (0.267) & (0.275) & (0.260) \\
Distance to nearest 2016 homicide (km) & & -0.156*** & -0.123*** & -0.131*** & -0.0894*** \\
 & & (0.0266) & (0.0247) & (0.0262) & (0.0238) \\
Homicide Count 2016 & & 0.464*** & 0.412*** & 0.452*** & 0.373*** \\
 & & (0.0868) & (0.0830) & (0.0846) & (0.0779) \\
asinh(311 Calls - NYPD) & & & & 0.371*** & 0.0731** \\
 & & & & (0.0245) & (0.0251) \\
asinh(311 Calls - HPD) & & & & & -0.0190 \\
 & & & & & (0.0119) \\
asinh(311 Calls - DOT) & & & & & 0.325*** \\
 & & & & & (0.0237) \\
asinh(311 Calls - DEP) & & & & & 0.0898*** \\
 & & & & & (0.0254) \\
asinh(311 Calls - DSNY) & & & & & -0.0614** \\
 & & & & & (0.0237) \\
asinh(311 Calls - DOB) & & & & & 0.0853*** \\
 & & & & & (0.0214) \\
asinh(311 Calls - DPR) & & & & & -0.138*** \\
 & & & & & (0.0186) \\
asinh(311 Calls - DOHMH) & & & & & 0.184*** \\
 & & & & & (0.0195) \\
asinh(311 Calls - DHS) & & & & & 0.300*** \\
 & & & & & (0.0149) \\
asinh(Total 311 Calls) & & & 0.716*** & & \\
 & & & (0.0321) & & \\
 & & & & & \\
Observations & 6,226 & 5,821 & 5,821 & 5,821 & 5,821 \\
R-squared & 0.003 & 0.080 & 0.171 & 0.119 & 0.288 \\ \hline
 \end{tabular}
}
 {\footnotesize 
 
 \textit{Notes: } This table presents OLS estimates of exposure disparity among census block groups (BGs) in NYC. Robust standard errors in parentheses: *** p$<$0.001, ** p$<$0.01, * p$<$0.05, + p$<$0.1.
}

\end{table}

\newpage
\begin{table}[H]
 \centering
 \caption{Disparities in Neighborhood Police Exposure (During Non-working Hours)}
 \label{tab:nonwork}
 \resizebox{\linewidth}{!}{%
 \begin{tabular}{lccccc} \hline
 & (1) & (2) & (3) & (4) & (5) \\
 VARIABLES & arsinh(Hours) & arsinh(Hours) & arsinh(Hours) & arsinh(Hours) & arsinh(Hours) \\ \hline
Relative Black Share & 0.0859*** & 0.0653*** & 0.0314*** & 0.0621*** & 0.0811*** \\
 & (0.00703) & (0.00784) & (0.00769) & (0.00826) & (0.00866) \\
Police: Relative Black & & & & 0.361*** & -0.613*** \\
 & & & & (0.0324) & (0.0755) \\
Relative Black Share X Police: Relative Black & & & & -0.0770*** & 0.0930+ \\
 & & & & (0.0227) & (0.0543) \\
Supervisor: Relative Black & & & & & 0.574*** \\
 & & & & & (0.0386) \\
Relative Black Share X Supervisor: Relative Black & & & & & -0.103*** \\
 & & & & & (0.0269) \\
Relative Hispanic Share & 0.0637*** & 0.0418*** & 0.00131 & 0.0244* & 0.0401*** \\
 & (0.00941) & (0.0110) & (0.00977) & (0.0111) & (0.0114) \\
Relative Asian Share & 0.0165* & -0.0216*** & -0.0193*** & -0.0217*** & -0.0120+ \\
 & (0.00691) & (0.00568) & (0.00539) & (0.00590) & (0.00630) \\
Log Population & & 0.511*** & 0.410*** & 0.514*** & 0.528*** \\
 & & (0.0201) & (0.0201) & (0.0208) & (0.0212) \\
\% College Graduates & & 1.011*** & 0.813*** & 0.978*** & 1.151*** \\
 & & (0.0584) & (0.0571) & (0.0598) & (0.0609) \\
Median Household Income (1K) & & -0.00351*** & -0.00392*** & -0.00457*** & -0.00421*** \\
 & & (0.000370) & (0.000373) & (0.000382) & (0.000384) \\
Census Form Return Rate & & -0.849*** & -1.382*** & -0.575*** & -0.805*** \\
 & & (0.119) & (0.121) & (0.124) & (0.128) \\
Distance to nearest 2016 homicide (km) & & -0.0954*** & -0.113*** & -0.0754*** & -0.0836*** \\
 & & (0.00579) & (0.00617) & (0.00614) & (0.00676) \\
Homicide Count 2016 & & 0.183*** & 0.187*** & 0.198*** & 0.209*** \\
 & & (0.0199) & (0.0194) & (0.0204) & (0.0208) \\
 & & & & & \\
Observations & 23,682 & 22,521 & 22,521 & 20,961 & 20,112 \\
R-squared & 0.007 & 0.071 & 0.166 & 0.075 & 0.090 \\
 Fixed effects & NA & NA & City & NA & NA \\ \hline
 \end{tabular}
 }
 {\footnotesize 
 \textit{Notes: } This table presents OLS estimates of exposure disparity among census block groups (BGs) across 21 cities (Column 1,2,4,5) and within cities (Column 3). All race variables are mean-centered, relative Black (Hispanic, Asian) shares defined as ratio of \% Black (Hispanic, Asian) in BG to \% in city, Police (Supervisor) Relative Black defined as ratio of \% Black of sworn officers (supervisors) to \% in city. The dependent variable is police hours observed in BGs (during non-working hours), transformed in arsinh values. Household income is measured in thousands of dollars, census return rates range from 0-1. Robust standard errors are reported in parentheses: *** p$<$0.001, ** p$<$0.01, * p$<$0.05, + p$<$0.1.
}
\end{table}

\newpage
\begin{table}[H]
\centering
\caption{Disparities in NYC Neighborhood Police Exposure}
\label{tab:NYC}
\resizebox{\linewidth}{!}{%
\begin{tabular}{lcccc} \hline
 & (1) & (2) & (3) & (4) \\
 VARIABLES & \multicolumn{4}{c}{Police Exposure in a Census Block Group: arsinh(Hours)} \\ \hline
 & & & & \\
 Relative Black Share & 0.0427** & 0.127*** & 0.0567*** & 0.119*** \\
 & (0.0147) & (0.0185) & (0.0147) & (0.0185) \\
Relative Hispanic Share & 0.0375+ & 0.212*** & 0.0536** & 0.180*** \\
 & (0.0207) & (0.0309) & (0.0207) & (0.0307) \\
Relative Asian Share & -0.0198 & 0.0374* & -0.0162 & 0.0283+ \\
 & (0.0154) & (0.0169) & (0.0153) & (0.0167) \\
Log Population & & 0.455*** & & 0.428*** \\
 & & (0.0488) & & (0.0479) \\
\% College Graduates & & 1.707*** & & 1.567*** \\
 & & (0.136) & & (0.137) \\
Median Household Income (1K) & & -0.00150* & & -0.00240*** \\
 & & (0.000735) & & (0.000704) \\
Census Form Return Rate & & -0.322 & & -0.190 \\
 & & (0.250) & & (0.251) \\
Distance to nearest 2016 homicide (km) & & -0.156*** & & -0.161*** \\
 & & (0.0266) & & (0.0265) \\
Homicide Count 2016 & & 0.464*** & & 0.451*** \\
 & & (0.0868) & & (0.0857) \\
 & & & & \\
Observations & 6,226 & 5,821 & 6,062 & 5,672 \\
 R-squared & 0.003 & 0.080 & 0.005 & 0.073 \\ \hline
 \end{tabular}
}
 {\footnotesize 
 
 \textit{Notes: } This table presents the OLS regression estimates of the disparity in police presence among census block groups (BGs) in New York City. Column 1-2 includes the full sample; Column 3-4 excludes BGs in Precinct 1 (Wall Street), 6 (the West Village), 8 (Penn Station, Grand Central), 14 (Midtown South) and 18 (Midtown North). The dependent variable is the police hours observed in census block groups (excluding pings moving faster than 50 mph). Relative Black (Hispanic, Asian) shares are defined as the ratio of \% Black (Hispanic, Asian) in BG to \% Black (Hispanic, Asian) in the city. Household income is measured in thousands of dollars, census form return rates range from 0-1. Robust standard errors are reported in parentheses: *** p$<$0.001, ** p$<$0.01, * p$<$0.05, + p$<$0.1.
 }
\end{table}

\newpage
\begin{table}[H]
\centering
\caption{Disparities in Neighborhood Police Exposure and Downstream (Stop) Disparities}
\label{tab:stop}
\resizebox{\linewidth}{!}{
\begin{tabular}{lcccccc} \hline
 & (1) & (2) & (3) & (4) & (5) & (6) \\
VARIABLES & arsinh(Hours) & arsinh(Stops) & arsinh(Stops/Hours) & arsinh(Hours) & arsinh(Stops) & arsinh(Stops/Hours) \\ \hline
 & & & & & & \\
Relative Black Share & 0.0815*** & 0.259*** & 0.129*** & 0.0712*** & 0.157*** & 0.0723*** \\
 & (0.00996) & (0.0169) & (0.0137) & (0.0121) & (0.0186) & (0.0157) \\
Relative Hispanic Share & 0.0310** & 0.205*** & 0.112*** & 0.0390* & 0.0409 & 0.00932 \\
 & (0.0120) & (0.0219) & (0.0195) & (0.0152) & (0.0276) & (0.0247) \\
Relative Asian Share & 0.0262** & -0.0648*** & -0.0724*** & -0.0123 & -0.0491** & -0.0273* \\
 & (0.00916) & (0.0159) & (0.0124) & (0.00953) & (0.0158) & (0.0127) \\
Log Population & & & & 0.473*** & 0.136** & -0.242*** \\
 & & & & (0.0279) & (0.0422) & (0.0341) \\
\% College Graduates & & & & 1.116*** & 0.192 & -0.617*** \\
 & & & & (0.0786) & (0.124) & (0.102) \\
Median Household Income (1K) & & & & -0.00410*** & -0.0201*** & -0.0109*** \\
 & & & & (0.000526) & (0.000792) & (0.000615) \\
Census Form Return Rate & & & & -0.672*** & 6.476*** & 5.384*** \\
 & & & & (0.151) & (0.230) & (0.191) \\
Distance to nearest 2016 homicide (km) & & & & -0.0759*** & 0.143*** & 0.152*** \\
 & & & & (0.00892) & (0.0145) & (0.0131) \\
Homicide Count 2016 & & & & 0.226*** & 1.076*** & 0.677*** \\
 & & & & (0.0249) & (0.0385) & (0.0347) \\
 & & & & & & \\
Observations & 13,969 & 13,969 & 13,912 & 13,176 & 13,176 & 13,123 \\
 R-squared & 0.005 & 0.026 & 0.015 & 0.059 & 0.174 & 0.139 \\ \hline
\end{tabular}}

 {\footnotesize \textit{Notes: }All race variables are mean-centered, relative Black (Hispanic, Asian) shares are defined as the ratio of \% Black (Hispanic, Asian) in a BG to the \% in that city, Police (Supervisor) Relative Black defined as the ratio of \% Black of a department's sworn officers (supervisors) to the \% in that city. Household income is measured in thousands of dollars, census return rates range from 0-1. Robust standard errors are reported in parentheses: *** p$<$0.001, ** p$<$0.01, * p$<$0.05, + p$<$0.1.}
\end{table}

\bibliographystyle{econ}
\bibliography{REFERENCES}